\documentclass{article}

\usepackage[english]{babel}

\usepackage[letterpaper,top=2cm,bottom=2cm,left=3cm,right=3cm,marginparwidth=1.75cm]{geometry}

\usepackage{amsmath}
\usepackage{amssymb}
\usepackage[pdftex]{graphicx}
\usepackage[colorlinks=true, allcolors=blue]{hyperref}
\usepackage{booktabs} 
\usepackage{mdframed}
\usepackage{authblk}

\usepackage{xcolor}

\title{Numerical explorations of solvent borne adhesives: A lattice-based approach to morphology formation}

\author[1]{Vì Cecilia Erik Kronberg}
\author[2]{Stela Andrea Muntean}
\author[3]{Nils Hendrik Kr\"oger}
\author[4]{Adrian Muntean}
\affil[1]{Department of Mathematics and Computer Science, Eindhoven University of Technology, Eindhoven, The Netherlands}
\affil[2]{Department of Engineering and Physics, Karlstad University, Karlstad, Sweden}
\affil[3]{tesa SE, Norderstedt, Germany}
\affil[4]{Department of Mathematics and Computer Science, Karlstad University, Karlstad, Sweden}
\begin{document}
\maketitle

\begin{abstract}
The internal structure of adhesive tapes determines the effective mechanical properties. This holds true especially for blended systems, here consisting of acrylate and rubber phases. In this note, we propose a lattice-based model to study numerically the formation of internal morphologies within a four-component mixture (of discrete particles) where the solvent components evaporate. Mimicking numerically the interaction between rubber, acrylate, and two different types of solvents, relevant for the technology of adhesive  tapes, we aim to obtain realistic distributions of rubber ball-shaped morphologies --- they play a key role in the overall functionality of  those special adhesives. Our model incorporates the evaporation of both solvents and allows for tuning the strength of two essentially different solvent-solute interactions and of the temperature of the system.
\end{abstract}

\section{Introduction}

\begin{figure}[ht!]
    \centering
    \includegraphics[width=0.6\linewidth]{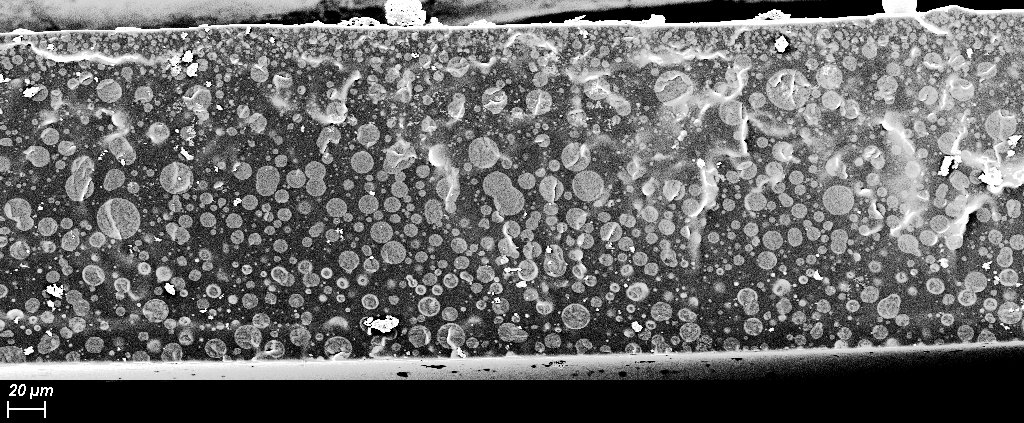}\hspace{10pt}%
    \includegraphics[width=0.33\linewidth]{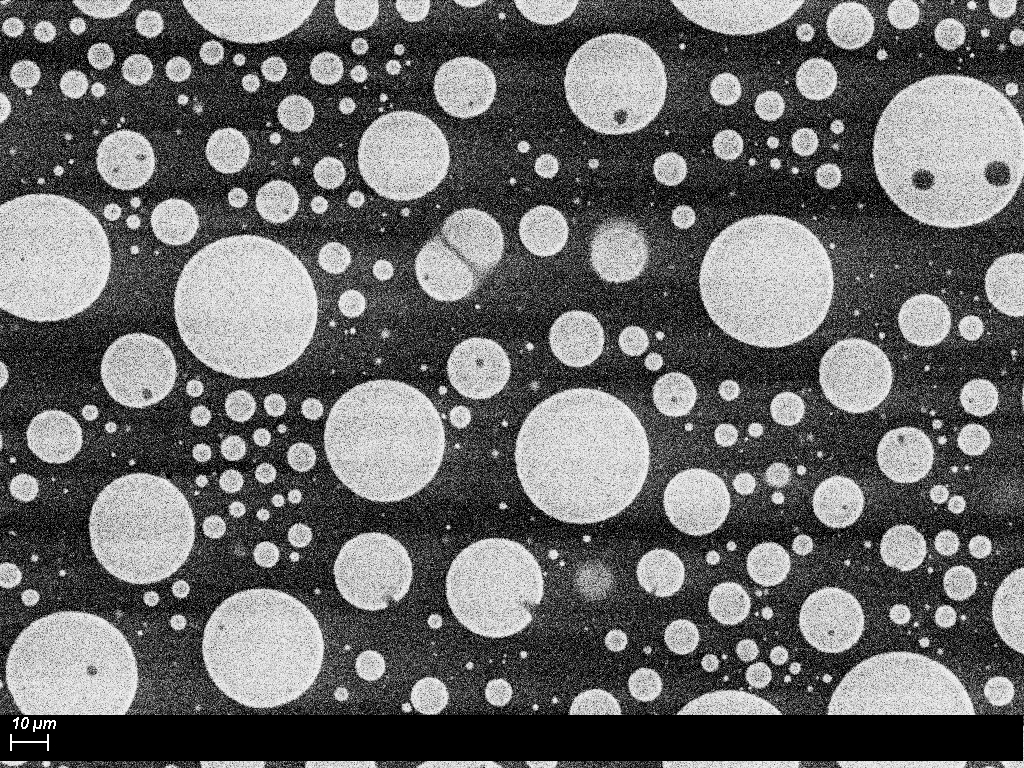}
    \caption{Left panel: Side view on the thin film. Right panel: Top view on the thin film. The ball-type structures are rubber parts;  within an acrylate matrix. The two solvents, ethylacetate and benzine, will evaporate fully, if the process is not stopped, and are indistinguishable in the experimental pictures. We see that multiple rubber disks coalesce to form larger ones, much like in our simulation.}
    \label{fig:exp}
 \end{figure}

An adhesive is usually a thin flexible layer that is applied on the boundary of objects aiming to join them together via an adhesive bonding process. The adhesive bonding is a complex process and its efficiency is directly influenced by the internal structure (here referred to as "morphology") of the layer. In this note, we focus our discussion on acrylic pressure sensitive adhesives (PSA), as they are one of the most important and widely used classes of adhesives. They have applications ranging from standard tapes and labels to special protective films (sealing, e.g., against oxidation or solar radiation). We refer the reader to \cite{Mapari} for a recent review of the classification of adhesive tapes and their applications. 
 One of many possibilities of enhancing the properties of PSAs is blending the acrylic base formulation with rubber. Here, our main interest lies in understanding the phase separation properties of a highly interacting multi-component mixture that is prepared in the production phase of solvent-borne adhesives. Specifically, and as a simplification, we investigate a combination of four species (acrylate, rubber, and two distinct solvents), allowing for the possibility of evaporating the two solvents.\footnote{In industrial formulations additional ingredients are added like  tackifying resins, wetting agents, anti-aging agents, plasticizers and more.}
A typical experimental picture we have in mind for this setting is shown in Figure \ref{fig:exp}. For this precise case, the two solvents are ethylacetate (solvent 1) and benzine (solvent 2), but the overall picture should be perceived in a generic way.

The main question we pose here is twofold: 
\begin{mdframed}[linecolor=black]
What influence does the temperature during the evaporation process have on the distribution of the rubber phase and on the dispersion of the two solvents  during the evaporation process?
\end{mdframed}

To address this question, we propose a study based on a stochastic lattice-type model that describes the initiation phase of the mixture, the actual dynamics leading to morphology formation (multi-component diffusion, interaction, and evaporation), and finally, a calibration phase --- referred here as migration phase. In this last phase, the obtained morphologies stabilize their terminal shape and any remaining solvent is allowed to disperse more evenly throughout the system with a switched-off evaporation process.
Within this frame, we apply the methodology that we developed for a numerical investigation of morphology formation as it occurs in the case of organic solar cells (OSC); see our recent work reported in \cite{Andrea_EPJ,Andrea_PhysRevE,Mario}.

The main common feature of both OSC and PSA systems is that the phase separation property arises during the interplay of the diffusion transport of a highly-interacting mixture of polymers and solvent, with the solvent evaporating until a certain mass fraction is reached. Obviously, the type of polymers and choices of solvents are very different in PSA compared to OSC, but the essential conceptual difference relies on the fact that for PSA systems, the temperature (and eventually also temperature gradients) play a prominent role in the formation of the final film and moreover, more solvents can be present in the mixture. Our description is stochastic, it holds at a discrete level (the lattice), and handles the time evolution and spatial localization of all the mass fractions in the mixture. On the other hand, as we are tracing only the evolution of mass fractions, no information about the transfer of momentum can be captured at this level. Consequently, neither fluid dynamics effects nor macroscopic mechanical responses of the film can be explored within our framework. To this end, conceptually different approaches need to be taken; we refer the reader to \cite{Silva} (adhesive materials and friction), \cite{Zeman} (dynamical adhesive contact in visco-elastic materials) and \cite{Bosco} (conservation of historical heritage) for remotely related situations where linear momentum information is handled computationally without aiming to capture insights in the dynamics of the phase separation. 

The paper is organized as follows. In Section \ref{Model}, we describe the lattice model, and in Section \ref{Results}, we show our simulation results and then discuss their relevance. The conclusion of this study as well as an outlook of further possible questions to be investigated in this setting are the subject of Section \ref{Outlook}. 

\section{The lattice-based model}\label{Model}

In this section, we give the details of our lattice-based model and describe the algorithm used to perform the simulations. The implementation of the algorithm is done in MATLAB and is publicly available at \url{github.com/vcekron/solventAdhesive}. 

\subsection{The lattice}
Consider a rectangular two-dimensional lattice $\Lambda = \{ 1, \dots, L_1\} \times \{1, \dots, L_2\}$, where $L_1 \geq L_2$.
An element of $\Lambda$ is called a \textit{site}.
Associated with each site are two \textit{bonds} --- one horizontal and one vertical --- connecting each site to two neighbouring sites.
The sites are populated by a species variable $\sigma \in \{1,2,3,4\}$, where the meaning of each species variable is explained in Table \ref{tab:colours}, where the ``color'' property refers to the colors in the figures presented in the results section; see Section~\ref{Results}.

\begin{table}[ht!]
    \caption{List of the mixture components with their coloring.}
    \label{tab:colours}
    \centering
    \begin{tabular}{ c l l }
        \toprule
        Species & Component & Color \\ [0.5ex] 
        \midrule
        1 & Acrylate & Blue \\ 
        2 & Rubber & Yellow \\
        3 & Ethylacetate & Red\\
        4 & Benzine & Green\\
        \bottomrule
    \end{tabular} 
\end{table}

\subsection{The interaction matrix} 
The interaction between two neighbouring sites depends on the two species --- captured in the interaction matrix, denoted $J$,
 \begin{equation}
    J := 
    \begin{pmatrix}
        J_{11} & J_{12} & J_{13} & J_{14}\\
        J_{21} & J_{22} & J_{23} & J_{24}\\
        J_{31} & J_{32} & J_{33} & J_{34}\\
        J_{41} & J_{42} & J_{43} & J_{44}
    \end{pmatrix},
\end{equation}
which incorporates a few specific features regarding the way the components of the mixture interact with each other.

Besides the symmetry of the interaction matrix (compare \cite{Hansen}), we impose the following additional constraints on the entries of $J$, namely $J_{11}=J_{33}=J_{44}=0$ (no self interaction for the acrylate and the two solvents), $J_{22} \ll 0$ (strong self-attraction for the rubber), $J_{34}>0$ (the two solvents repel each other), $J_{24}>0$ (rubber slightly repels solvent 2), $J_{23}>0$ (rubber strongly repels solvent 1, which instead is less repelled by acrylate), and $J_{12} \gg 0$ (rubber and acrylate strongly repel each other).
Furthermore, we only look to the case $J_{13}<J_{23}$ (solvent 1 repels acrylate more than it repels rubber).
In this study, we have fixed the interaction matrix to be
\begin{equation}\label{eq:JMatrix}
    J :=
    \begin{pmatrix}
        0 & 6 & 0.5 & 1.5\\
        6 & -4 & 1 & 0.5\\
        0.5 & 1 & 0 & 0.75\\
        1.5 & 0.5 & 0.75 & 0
    \end{pmatrix},
\end{equation}
in accordance with the description above.

In order to simulate the dynamics of the lattice system, let $\mathcal{H}$ represent the Hamiltonian
\begin{equation}
    \mathcal{H} := \sum_{\langle i,j \rangle} J_{\sigma_i,\sigma_j},
\end{equation}
where the sum is carried out over nearest-neighbours, and $J_{\sigma_i,\sigma_j}$ represents the components of $J$ in Eq.~\eqref{eq:JMatrix}.
The energy difference between two configurations is then
\begin{equation}\label{eq:deltaE}
    \Delta E := \mathcal{H}^\mathrm{proposed} - \mathcal{H}^\mathrm{current}.
\end{equation}
Here, the upper index "current" of $\mathcal{H}$ refers to the evaluation of the Hamiltonian on the current configuration, while the upper index "proposed" points out the evaluation of the Hamiltonian for the next configuration, where a particle wants to move to another site. 
Furthermore, let $\beta > 0$ be the inverse temperature, i.e., smaller $\beta$ is associated with a higher temperature, and {\em vice versa}. The precise choice of $\beta$ is essential in the fifth step of the Metropolis algorithm explained next.

\subsection{The Metropolis algorithm}
\noindent The Metropolis algorithm used in the simulations presented in the next section can be explained in five steps:
\begin{enumerate}
    \item pseudo-randomly select a site on $\Lambda$,
    \item pseudo-randomly select a bond associated with the site (vertical or horisontal),
    \item propose to switch the spins occupying the bonded sites,
    \item evaluate the energy difference, $\Delta E$ --- Eq.~\eqref{eq:deltaE}, associated with the switch,
    \item accept the proposed move with probability 1 if $\Delta E < 0$ and $\exp(-\beta \Delta E)$ otherwise.
\end{enumerate}
It is not difficult to see that higher temperature, i.e., smaller $\beta$ will lead to more proposed moves being accepted in the Metropolis algorithm.\\

It is beyond the scope of this work  to enter here into the technical details behind the Monte Carlo method and the Metropolis algorithm. Instead, we refer the reader, for instance, to \cite{Okabe,Santen} or to the textbook \cite{Ritter}.

\section{Simulation results}\label{Results}
This section shows the time evolution of the systems with a focus on two effects: the addition of the so-called disks formation stage and varying the system temperature.
For further insight into the dynamics of the system, please see the movies of these simulations publicly available at \url{github.com/vcekron/solventAdhesive}.

\subsection{The disks formation stage}
\begin{figure}[ht!]
    \centering
    \includegraphics[width=0.33\linewidth]{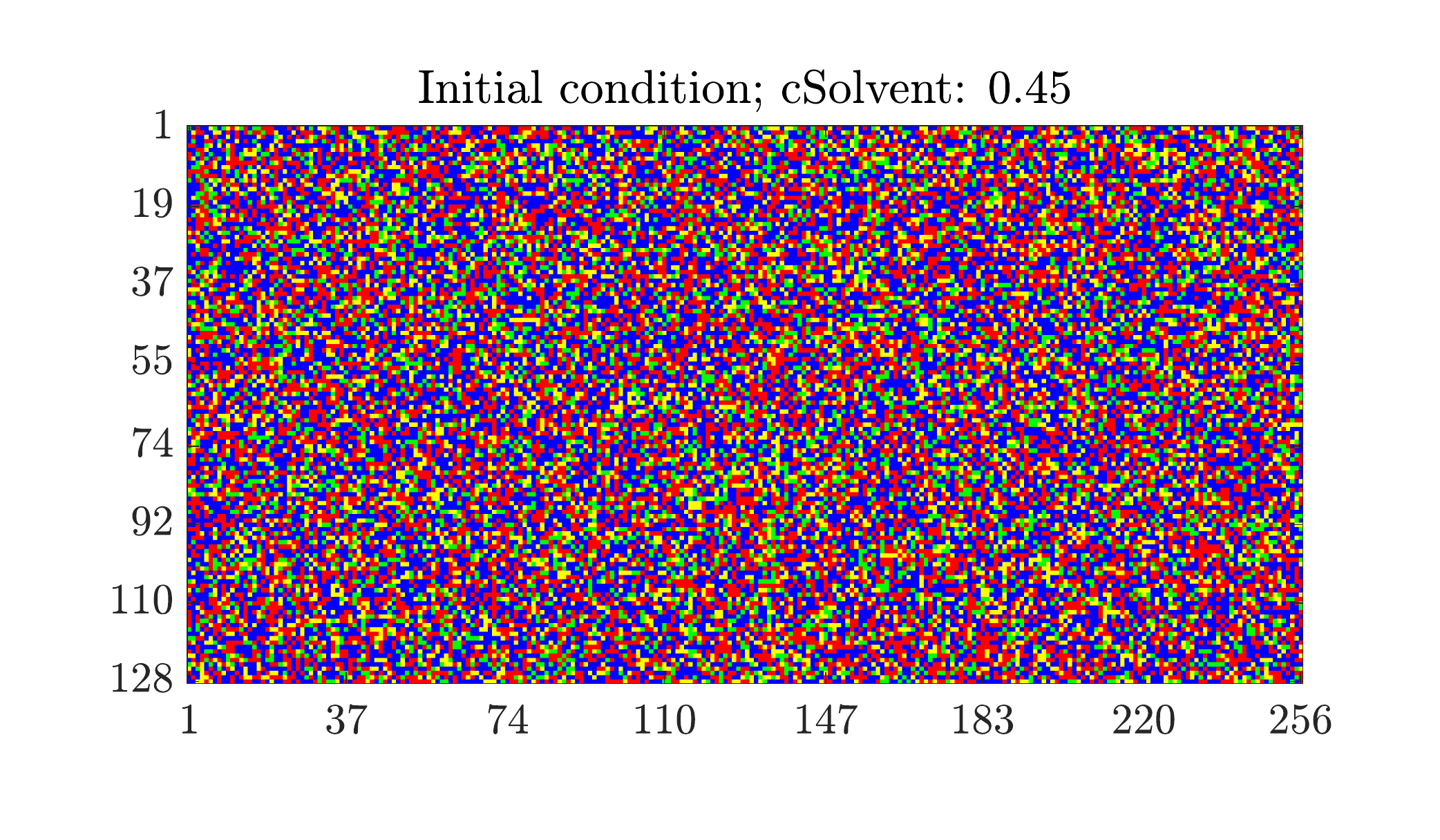}\\
    \hspace{0.33\linewidth}%
    \includegraphics[width=0.33\linewidth]{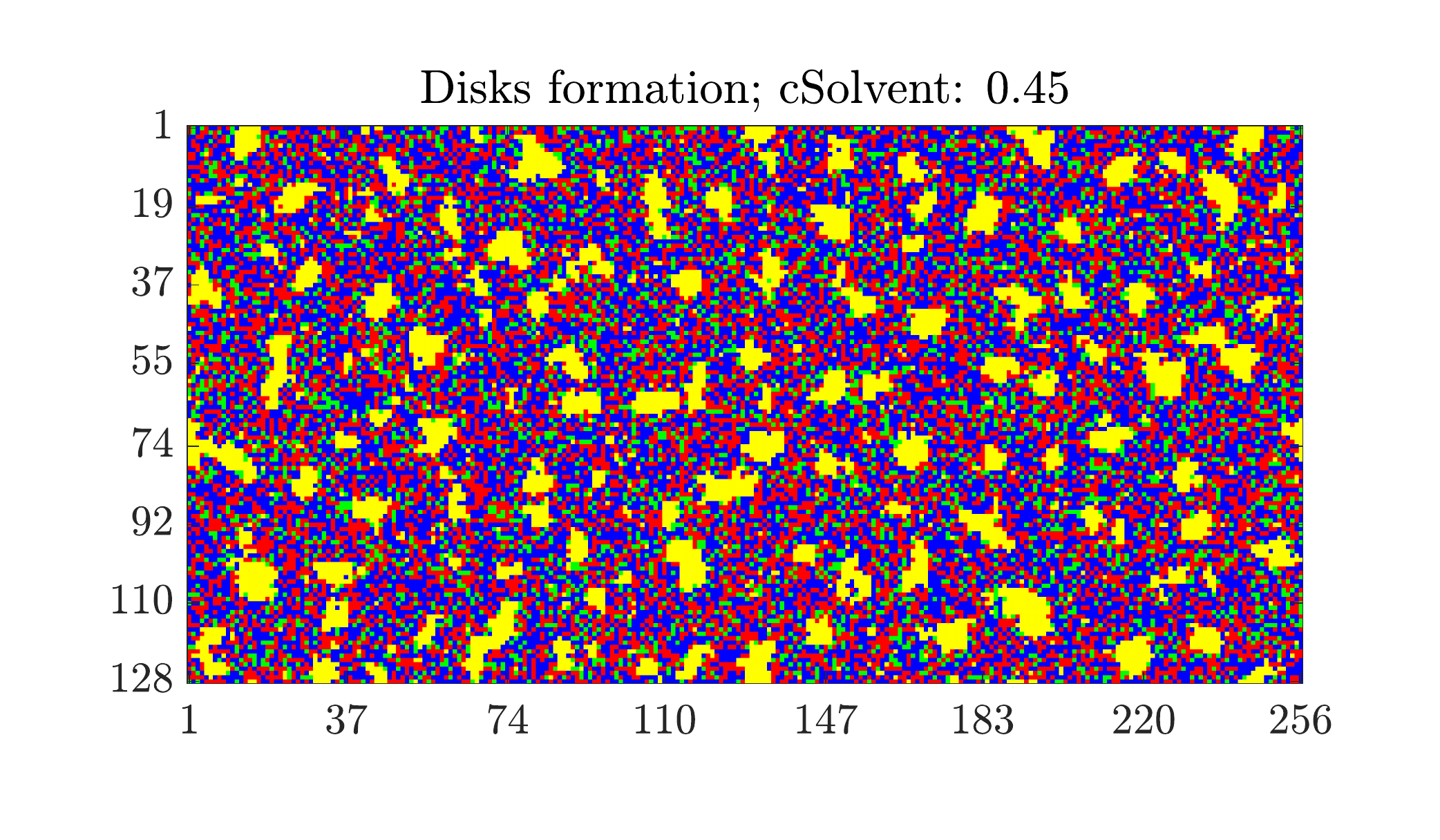}%
    \includegraphics[width=0.33\linewidth]{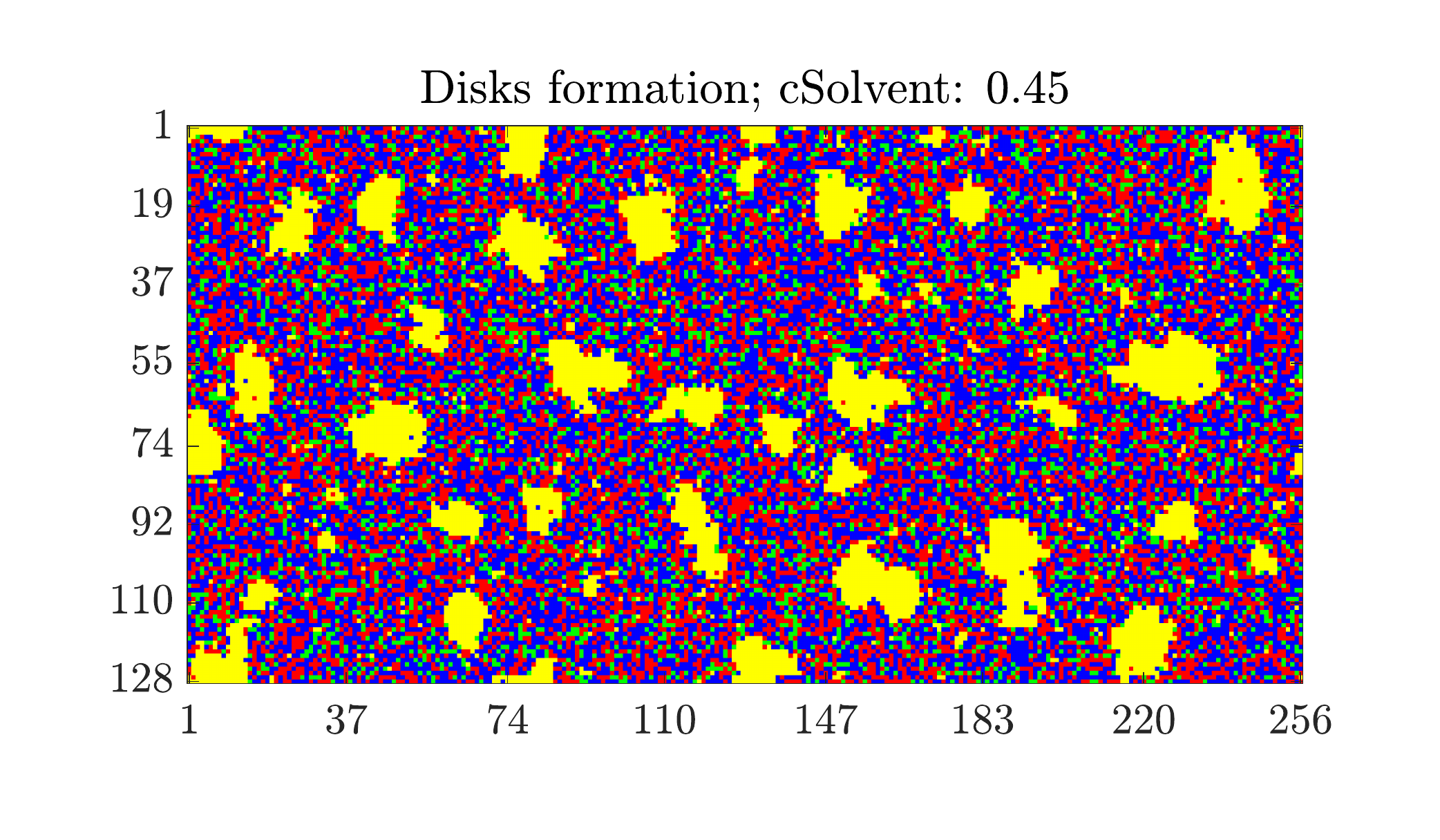}\\
    \includegraphics[width=0.33\linewidth]{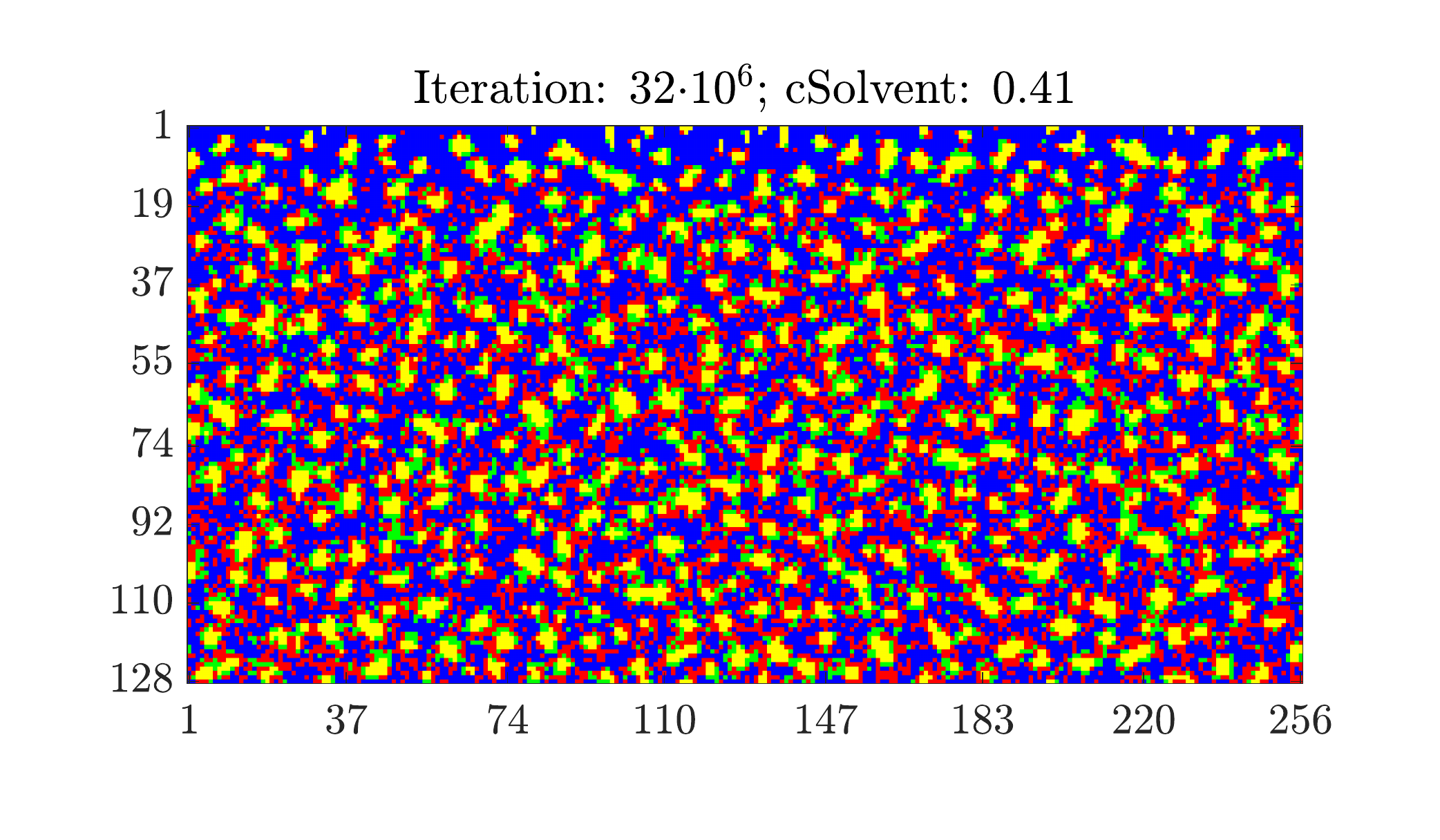}%
    \includegraphics[width=0.33\linewidth]{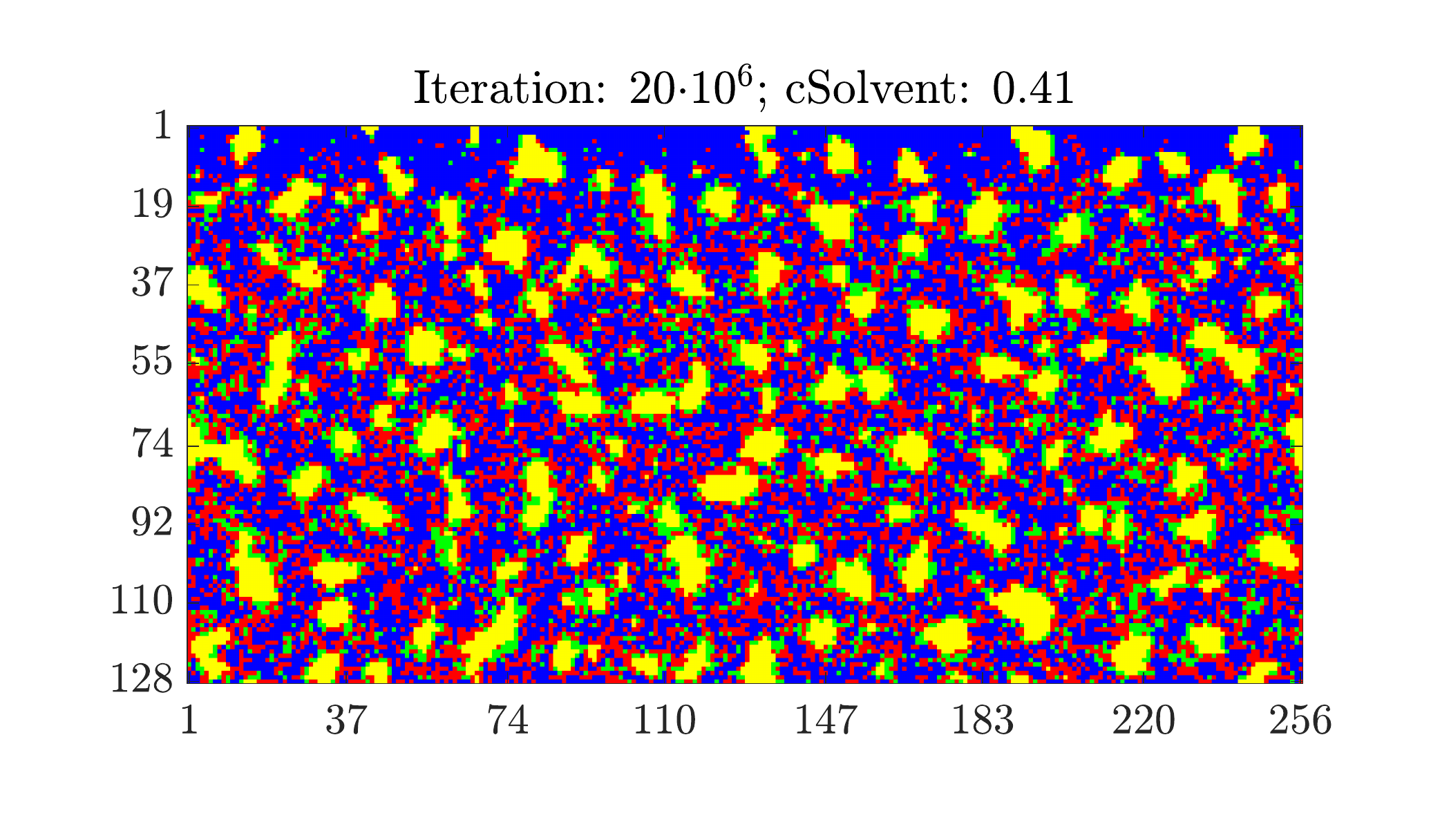}%
    \includegraphics[width=0.33\linewidth]{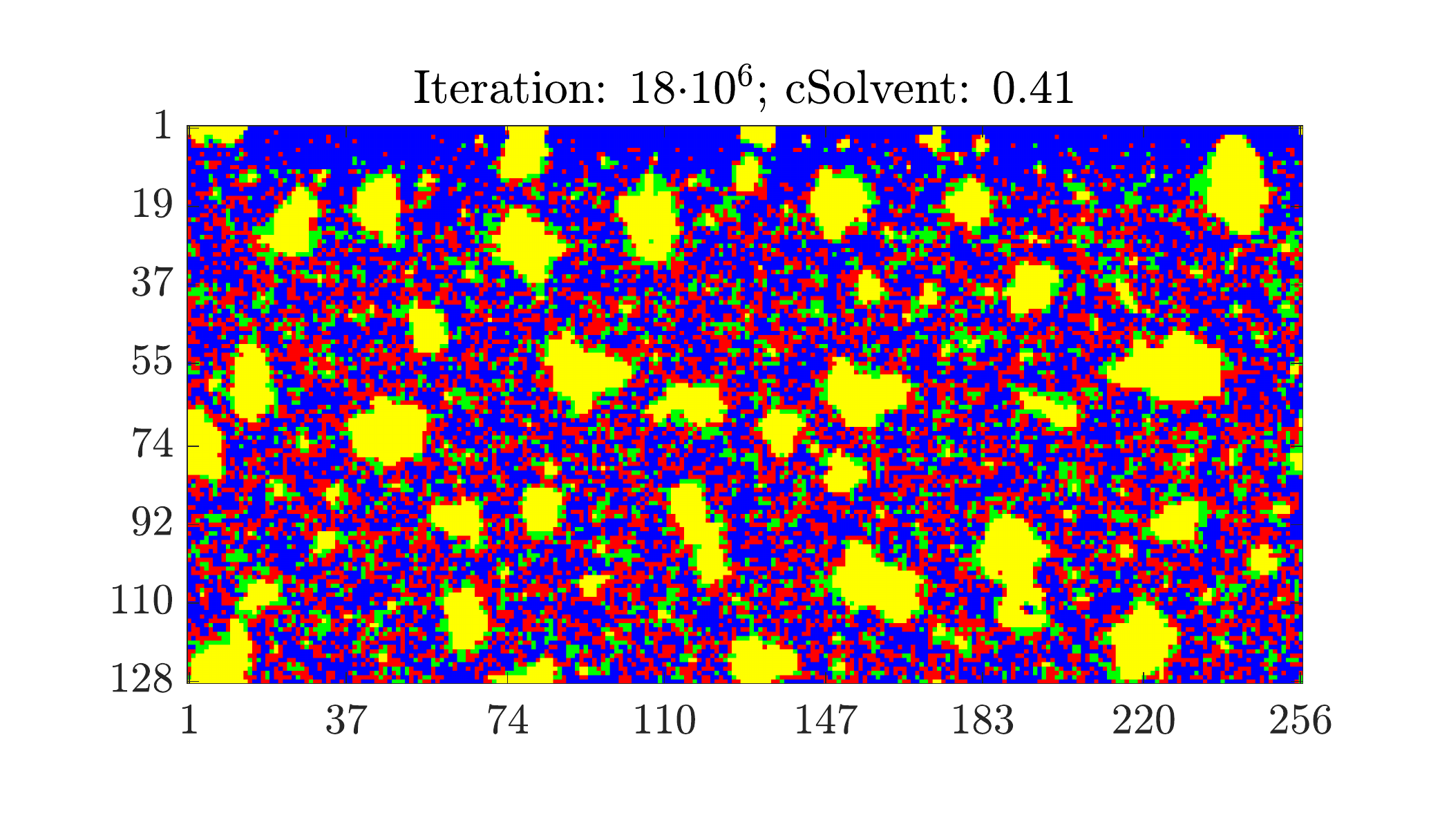}\\
    \includegraphics[width=0.33\linewidth]{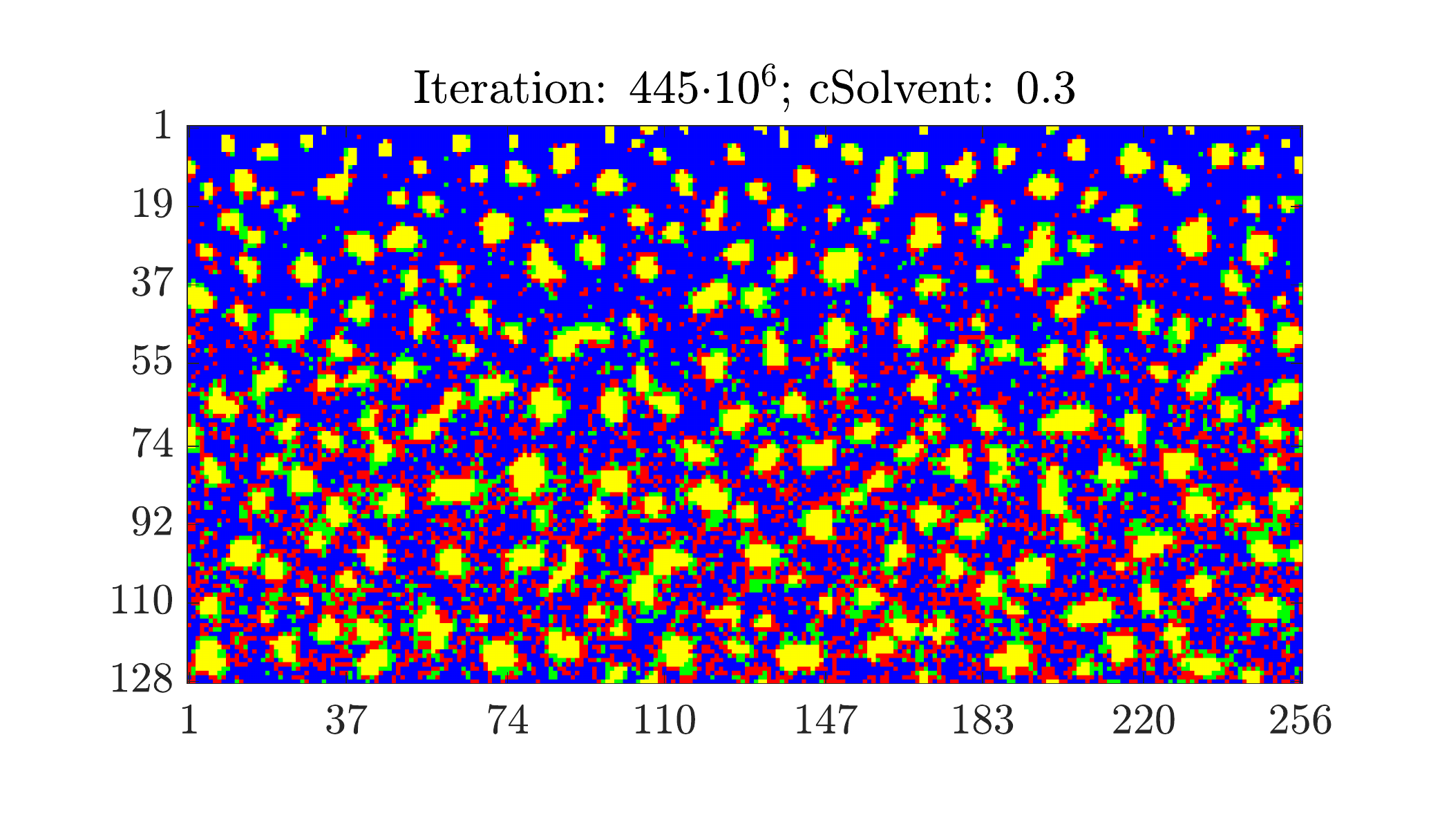}%
    \includegraphics[width=0.33\linewidth]{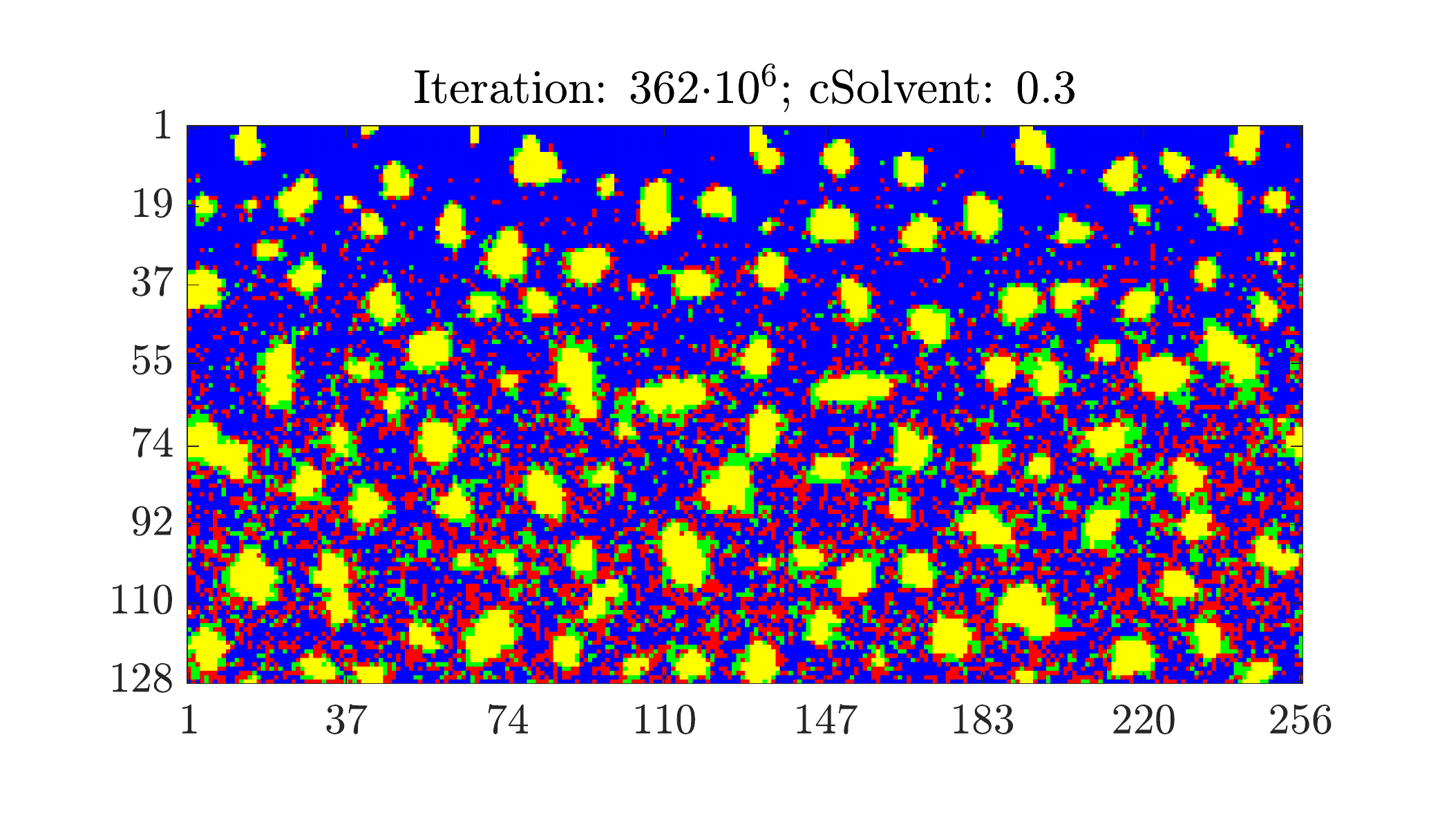}%
    \includegraphics[width=0.33\linewidth]{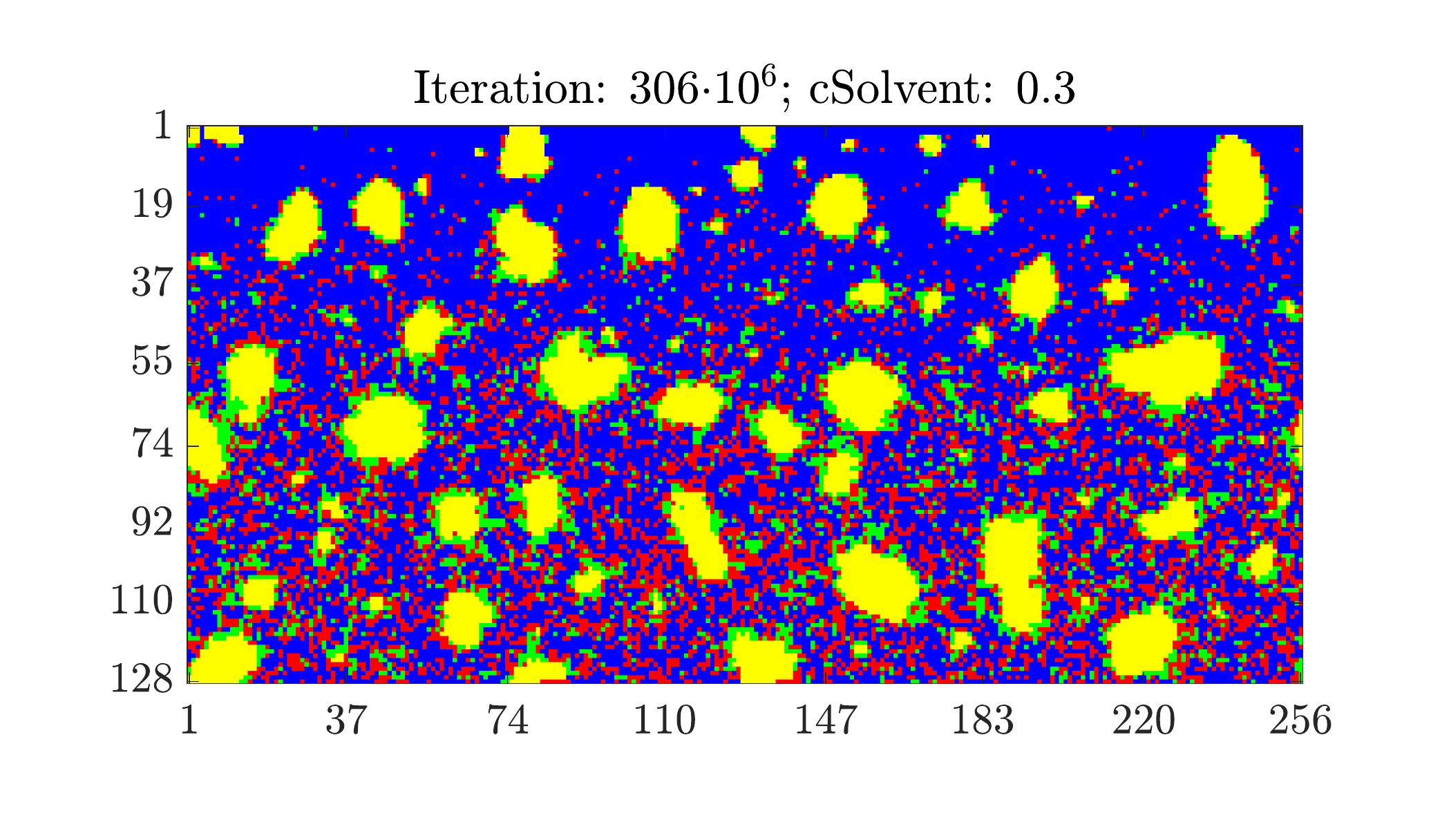}\\
    \includegraphics[width=0.33\linewidth]{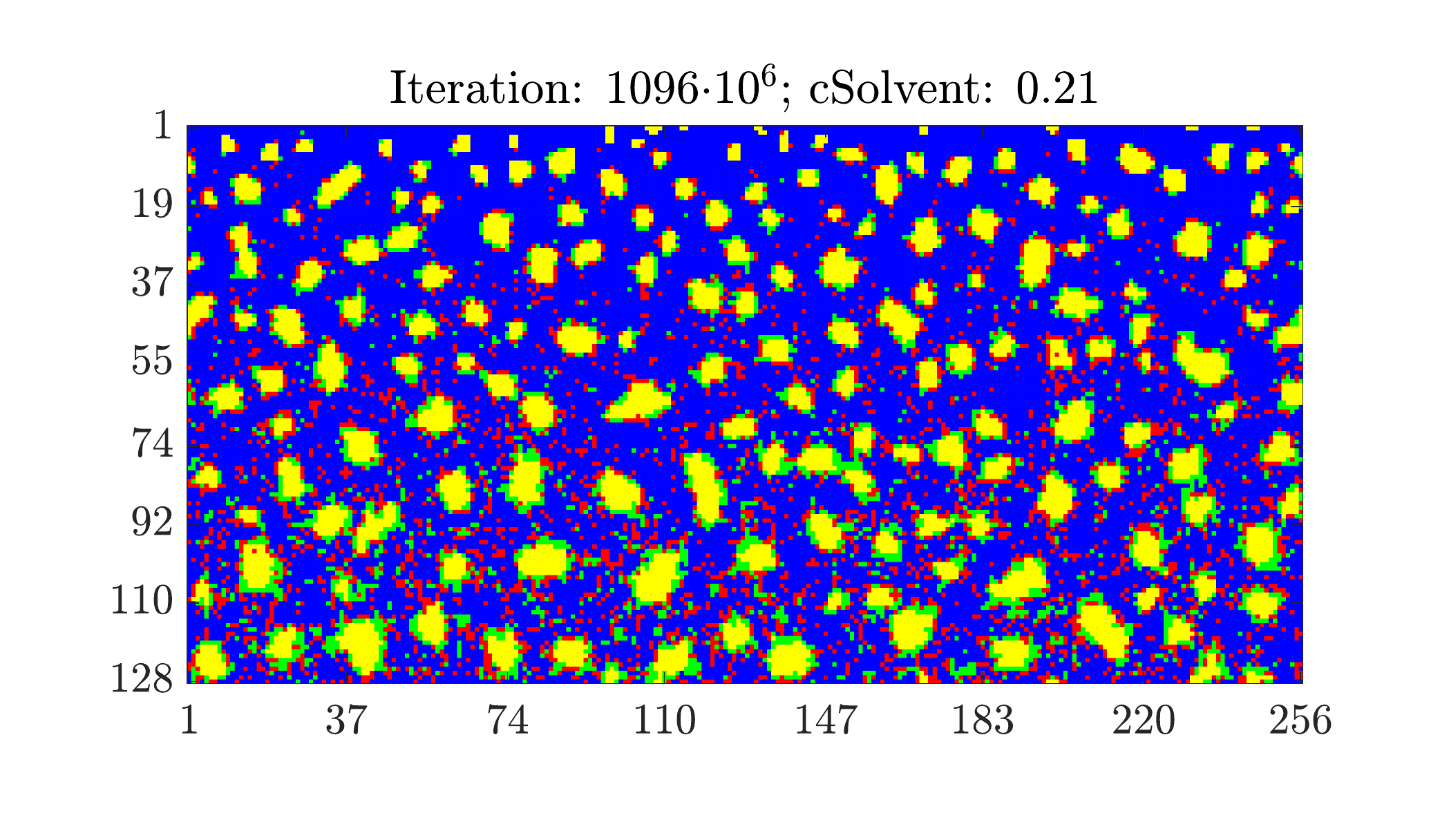}%
    \includegraphics[width=0.33\linewidth]{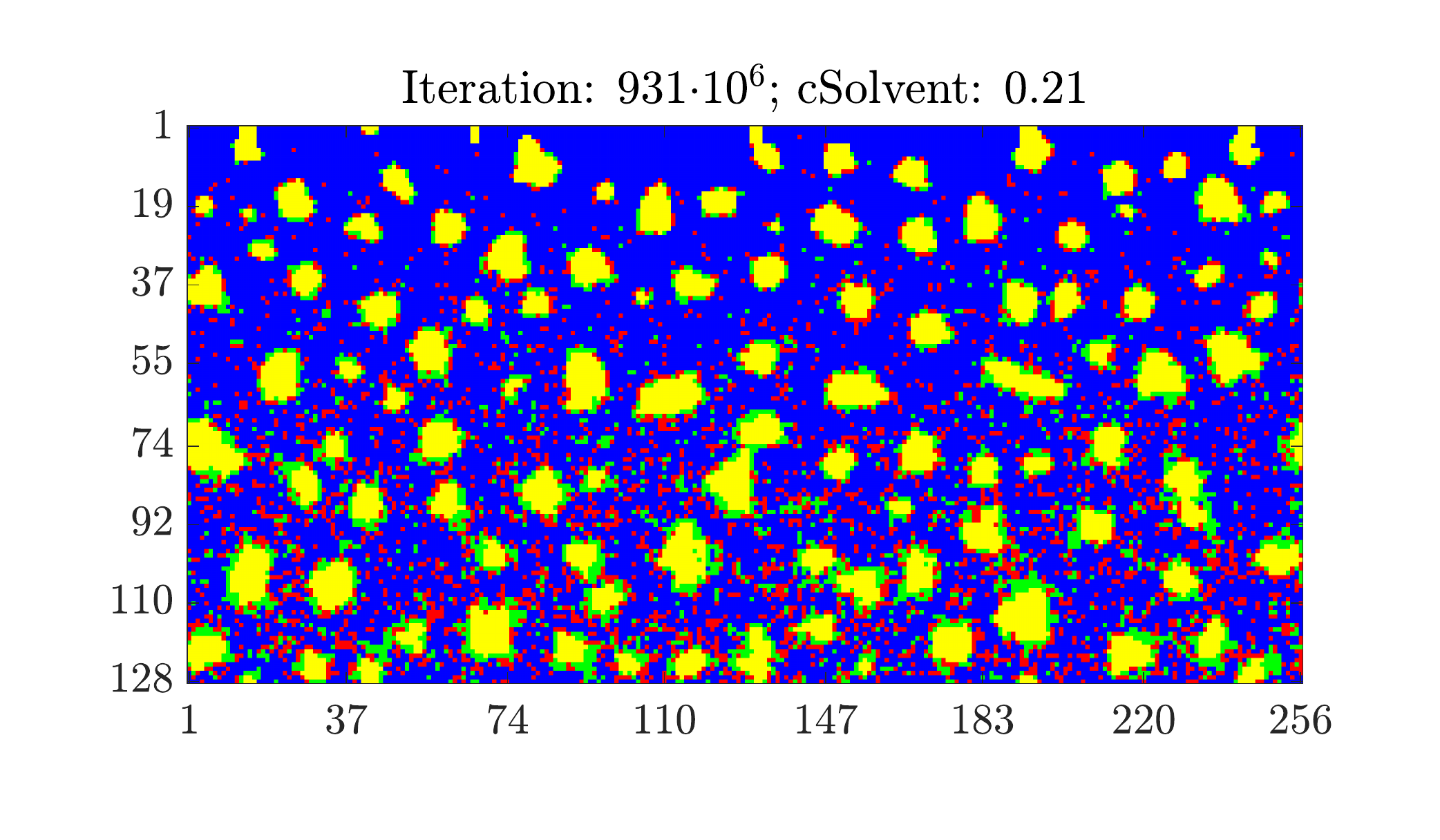}%
    \includegraphics[width=0.33\linewidth]{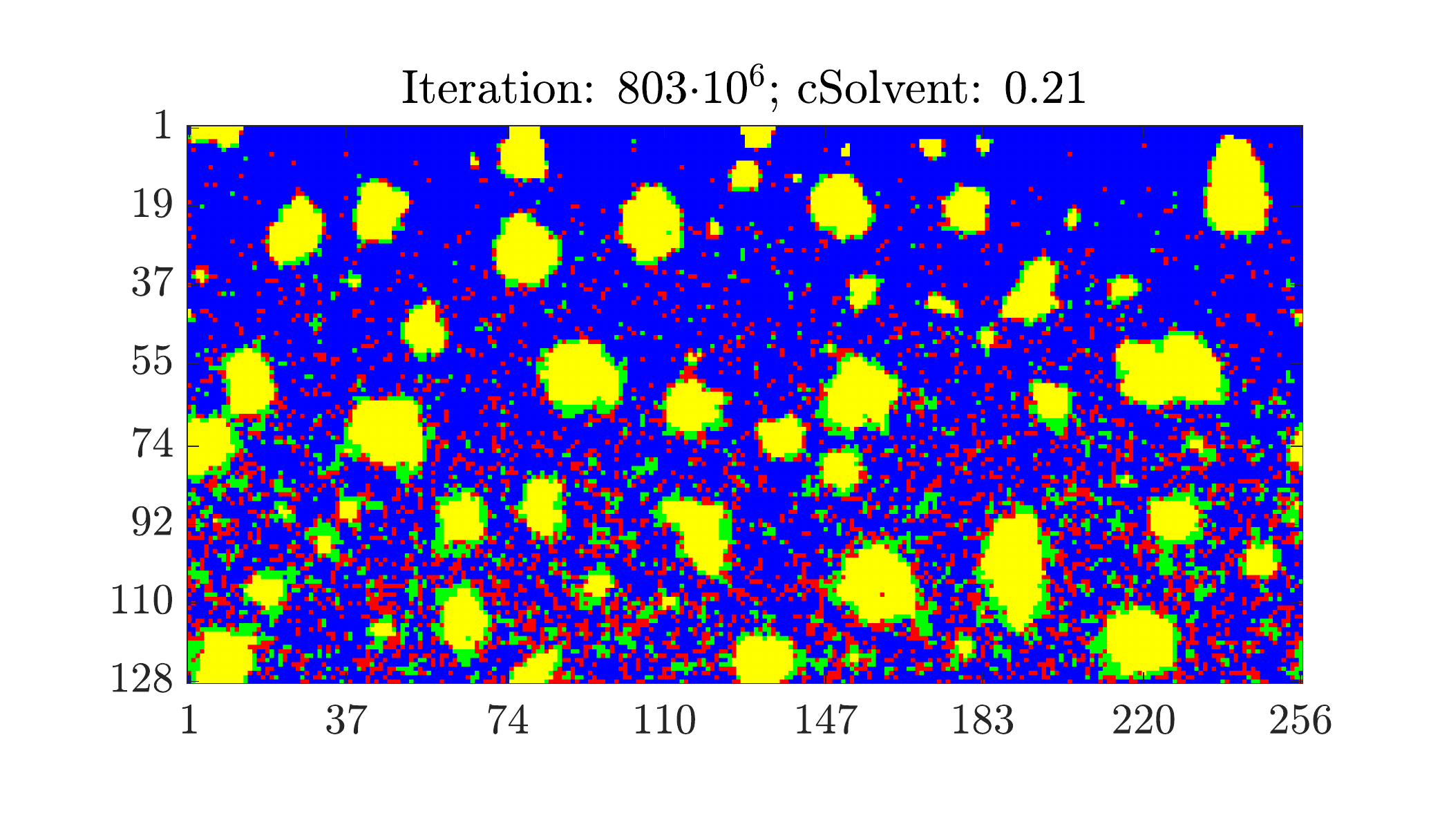}\\
    \includegraphics[width=0.33\linewidth]{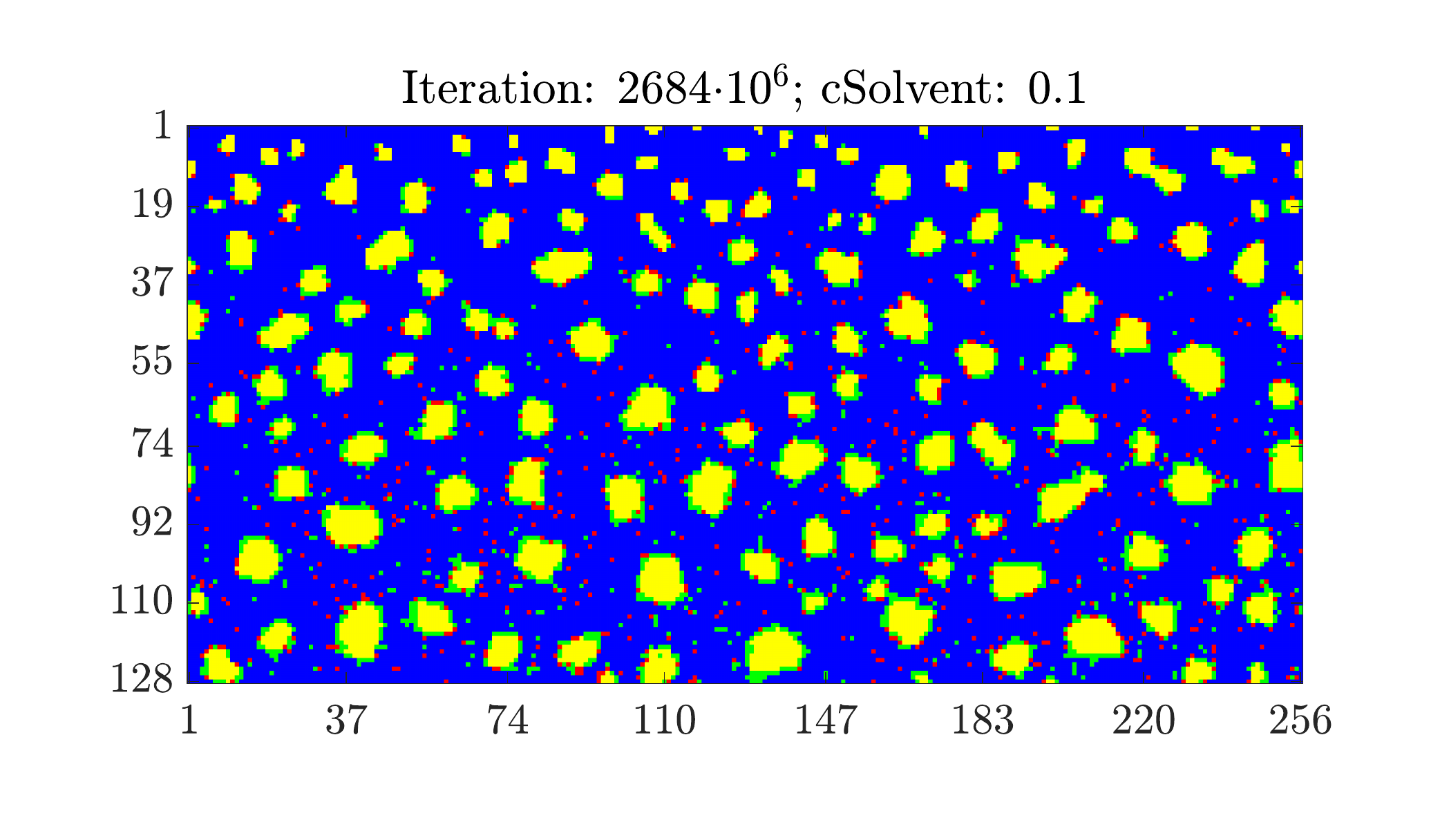}%
    \includegraphics[width=0.33\linewidth]{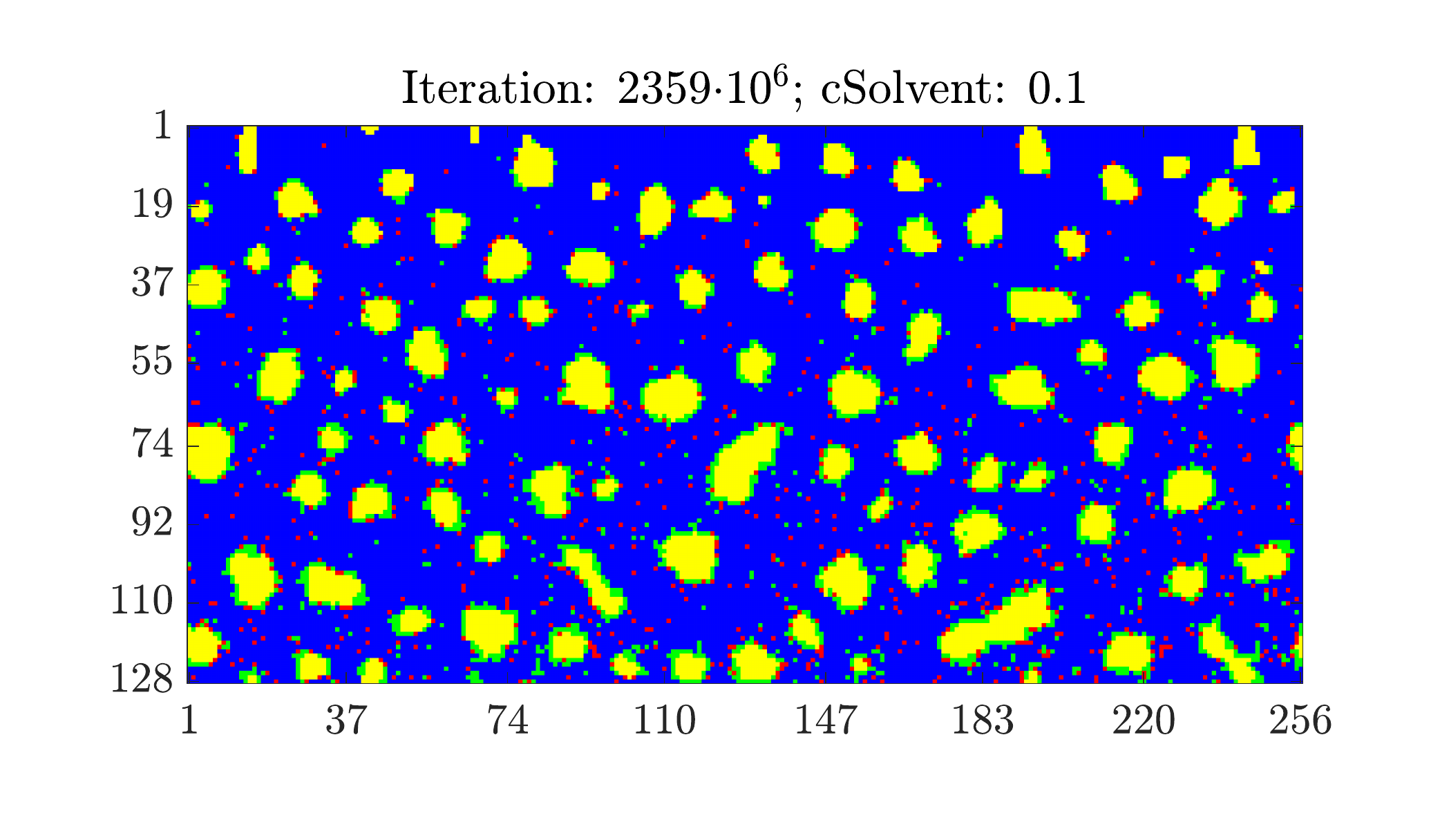}%
    \includegraphics[width=0.33\linewidth]{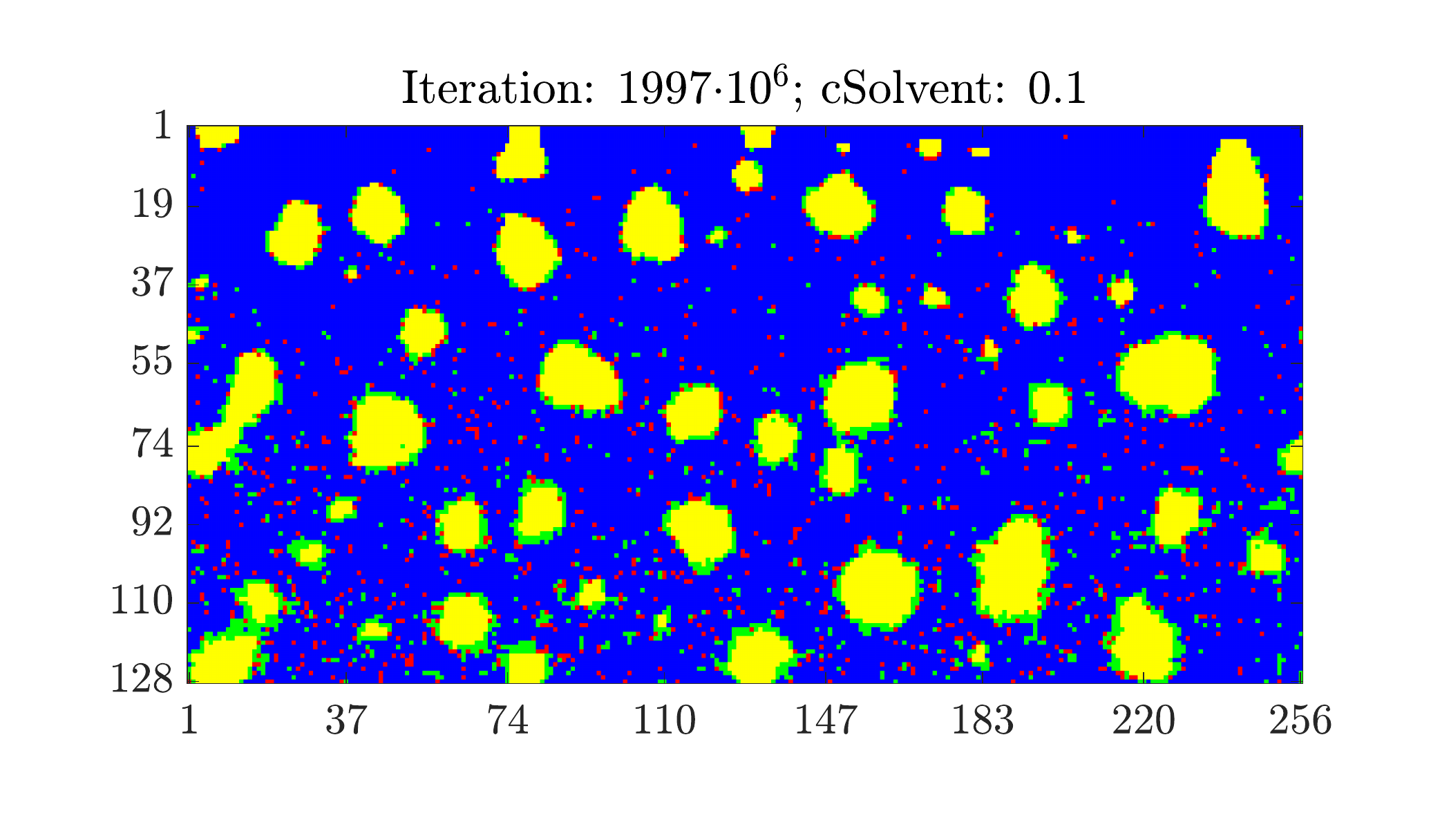}\\
    \includegraphics[width=0.33\linewidth]{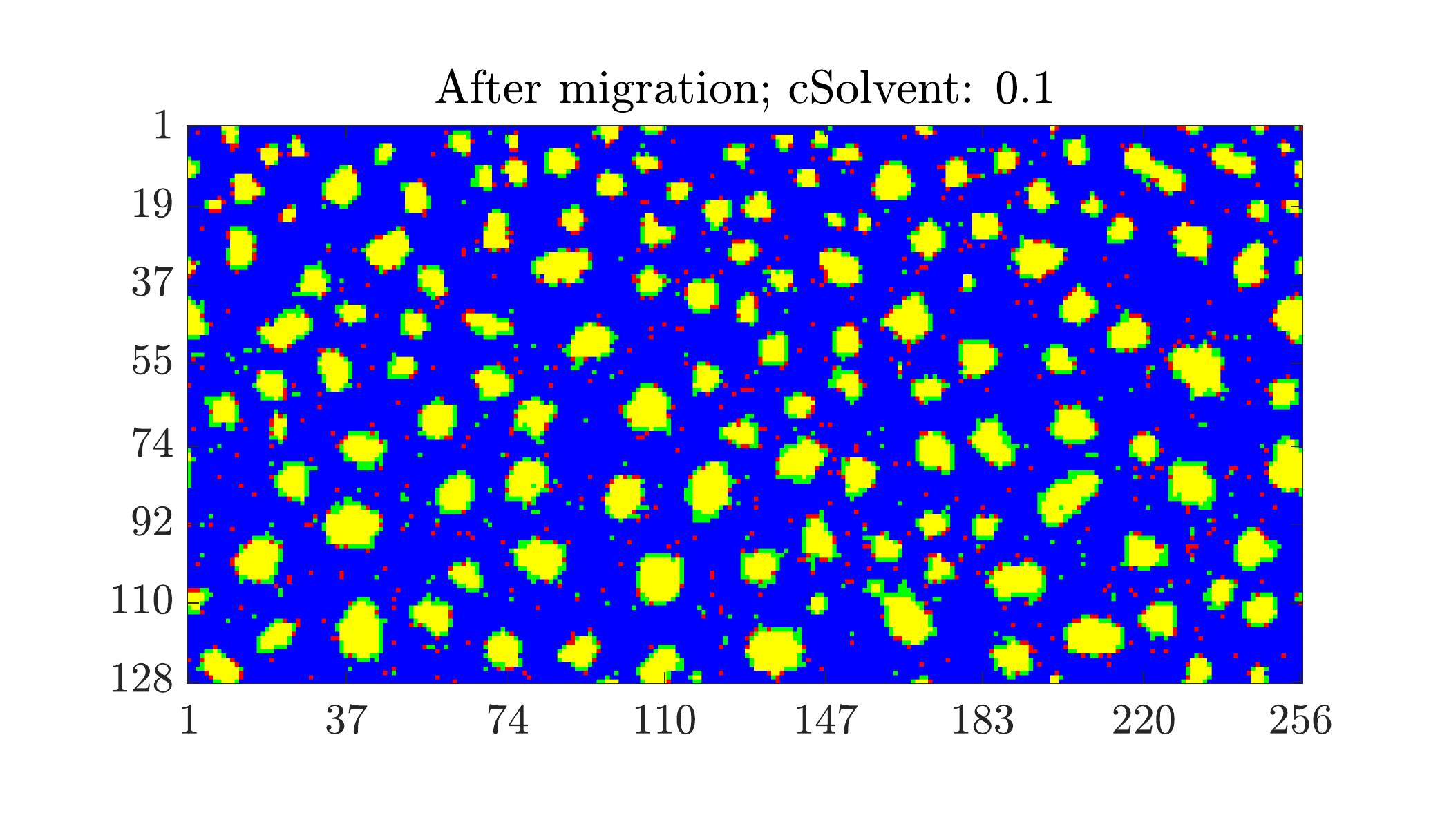}%
    \includegraphics[width=0.33\linewidth]{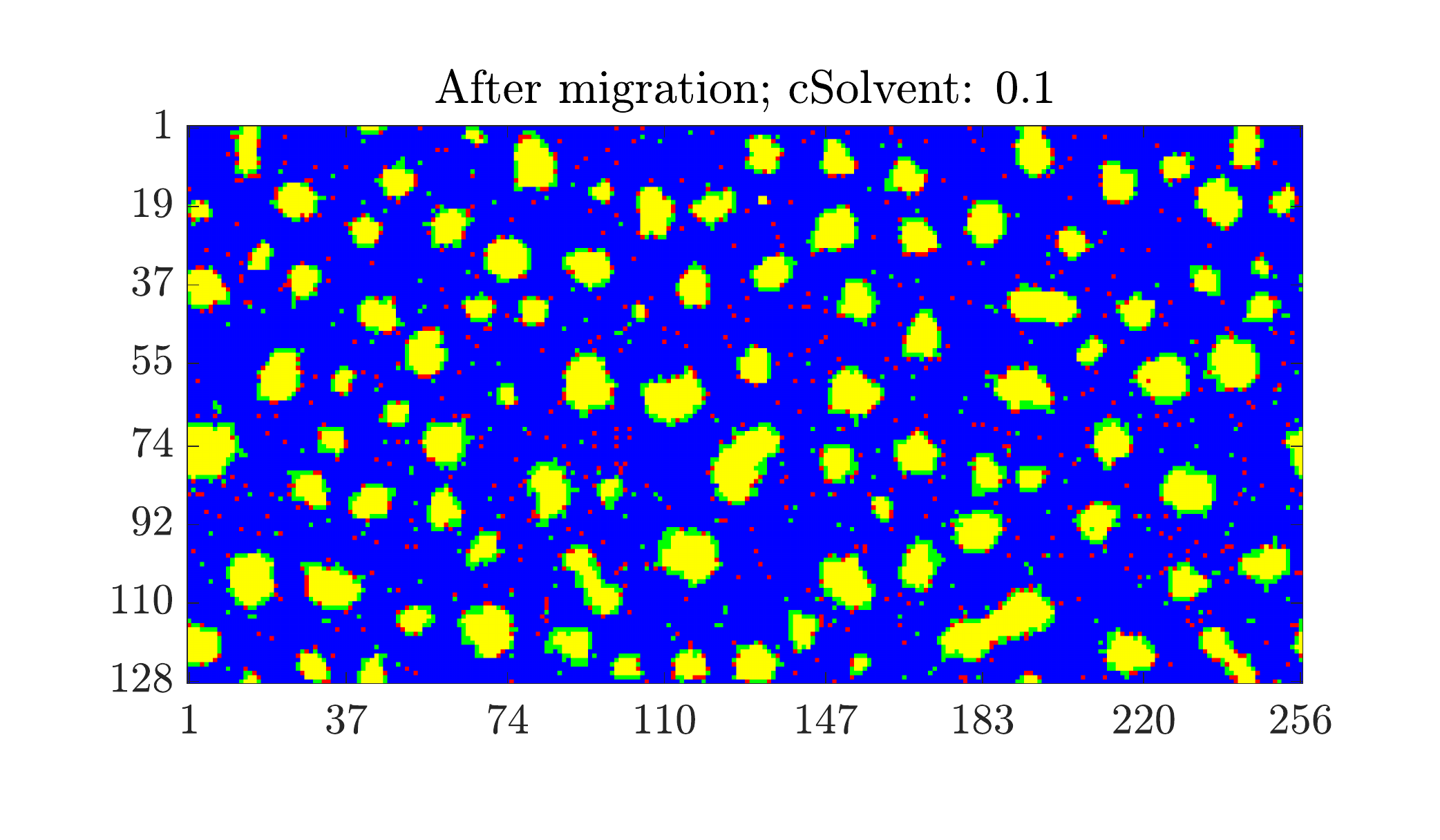}%
    \includegraphics[width=0.33\linewidth]{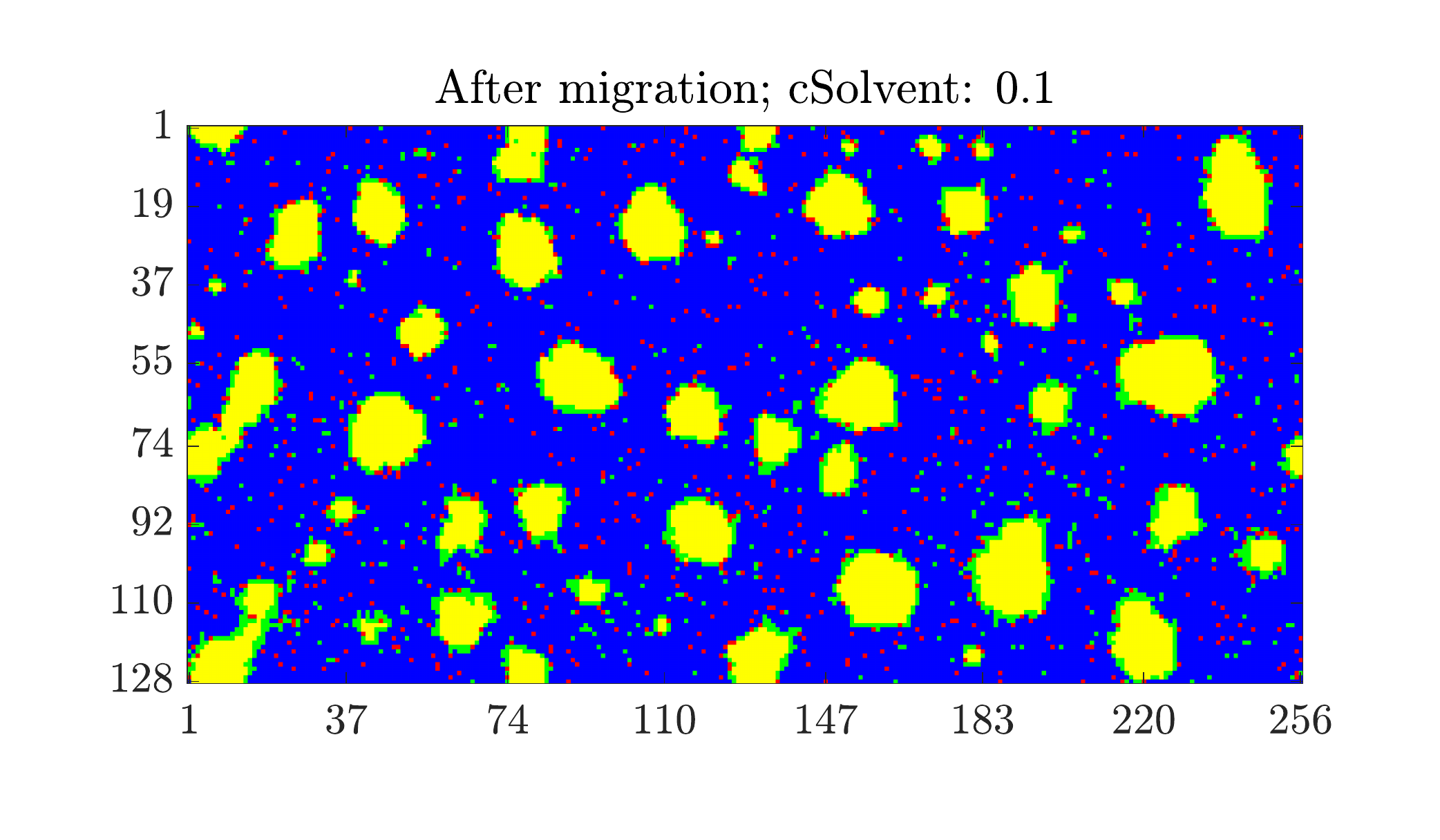}\\
    \caption{The effect of the disks formation stage at fixed temperature $\beta = 0.6$. Note that the disks formation stage has a drastic effect on the sizes of the yellow domains and their dispersity.}
    \label{fig:diskFormation}
\end{figure}

In order to study the effect of varying lengths of disk formation for fixed temperature, consider Fig.~\ref{fig:diskFormation}.
Here, the same initial system was used for all three cases (top row).
In the second row, the disks formation stage was carried out for zero iterations (first column), $10^8$ iterations (second column; approximately $3 \cdot 10^3$ Monte Carlo Steps (MCS)) and $10^9$ iterations (third column; approximately $3 \cdot 10^4$ MCS).
As was mentioned before, during the disks formation stage, only the rubber-rubber (yellow-yellow) interaction is switched on, and periodic boundary conditions are enforced everywhere without evaporation of the solvents.
That is,
\begin{equation}
    J_\mathrm{disks} = 
    \begin{pmatrix}
        0 & 0 & 0 & 0\\
        0 & -4 & 0 & 0\\
        0 & 0 & 0 & 0\\
        0 & 0 & 0 & 0
    \end{pmatrix}.
\end{equation}
Once the disks formation stage is over, the full dynamics, including solvents evaporation from the top row, are switched on and the system is allowed to evolve.
This evolution can be seen in consecutive rows after the second one, with snapshots at each 10\% reduction in the total amount of solvent present in the system.
Once the amount of solvent had reduced to 10\% of the lattice sites (second to last row), the full dynamics were switched off and the so-called migration stage started.
This stage lasted for $10^8$ iterations. During this stage, periodic boundary conditions are once again enforced such that the solvent can readily migrate throughout the domain.
The results of this stage is shown in the last row.

Focusing our attention at the last row, we can see that the disks formation stage has a drastic effect on the sizes of the yellow domains and their dispersity.
In particular, when the disks formation stage lasted for $10^9$ iterations (third column), the final domains were significantly larger compared to the shorter disks formation stage (second column).
Comparing the run without the disks formation stage to the middle one reveals less drastic changes in the end stage of the evolution, showing however clear differences in the initial stage, when there is significant amount of solvent in the system. 

The disks formation stage was introduced as a way to generate the rubber balls added to the mixture in the experimental setup\footnote{This stage can potentially be developed further by embedding an optimization step to search for optimal rubber distributions. This is though out of the scope of the current work. }; compare Figure \ref{fig:exp}. 
With this in mind, and comparing the results obtained herein, we have opted to run the remaining simulations with $10^8$ disks formation iterations as a compromise between omitting this stage and initialising with large yellow regions.

\begin{figure}[tbp]
    \centering
    \includegraphics[width=0.33\linewidth]{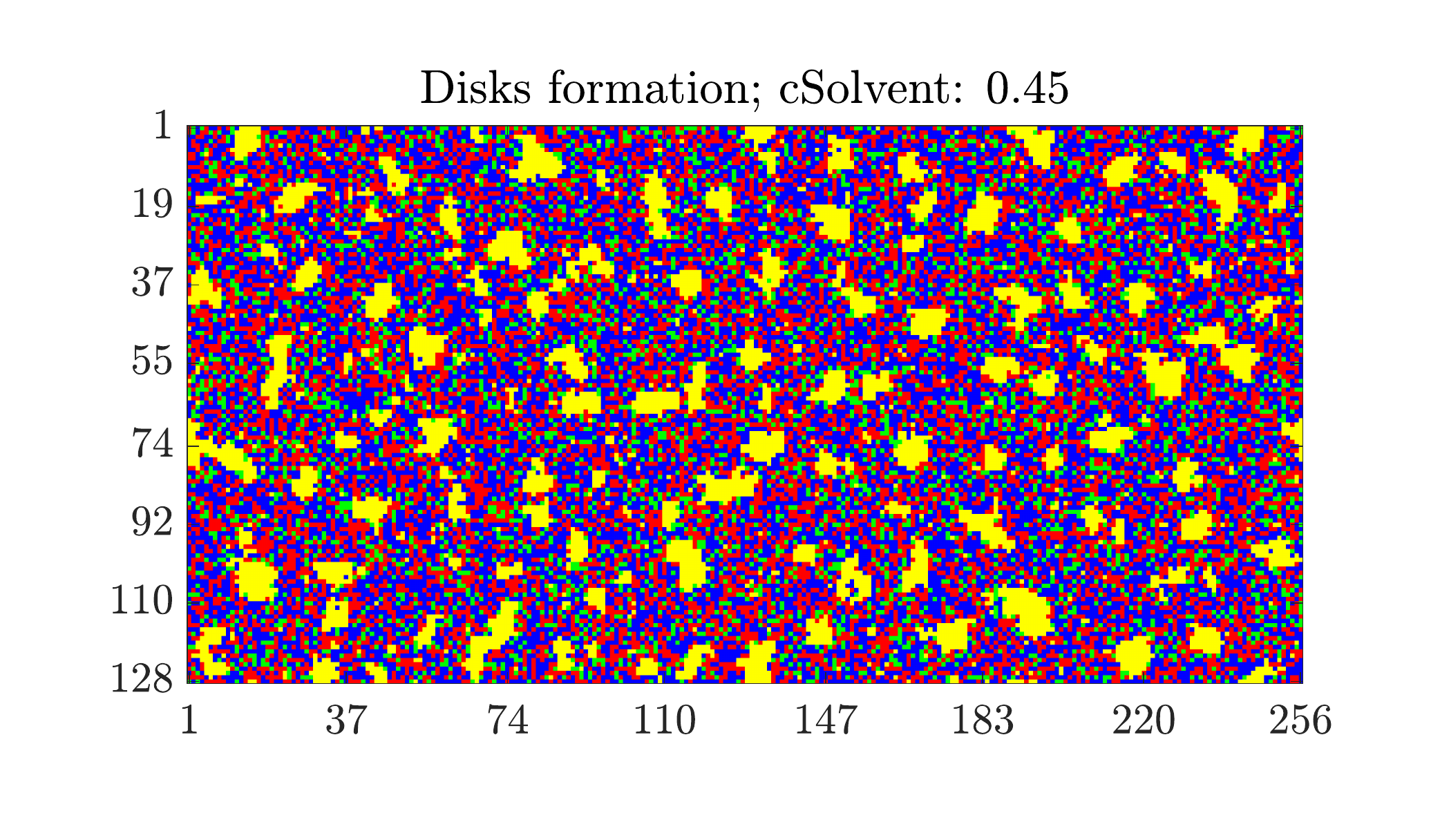}\\
    \includegraphics[width=0.33\linewidth]{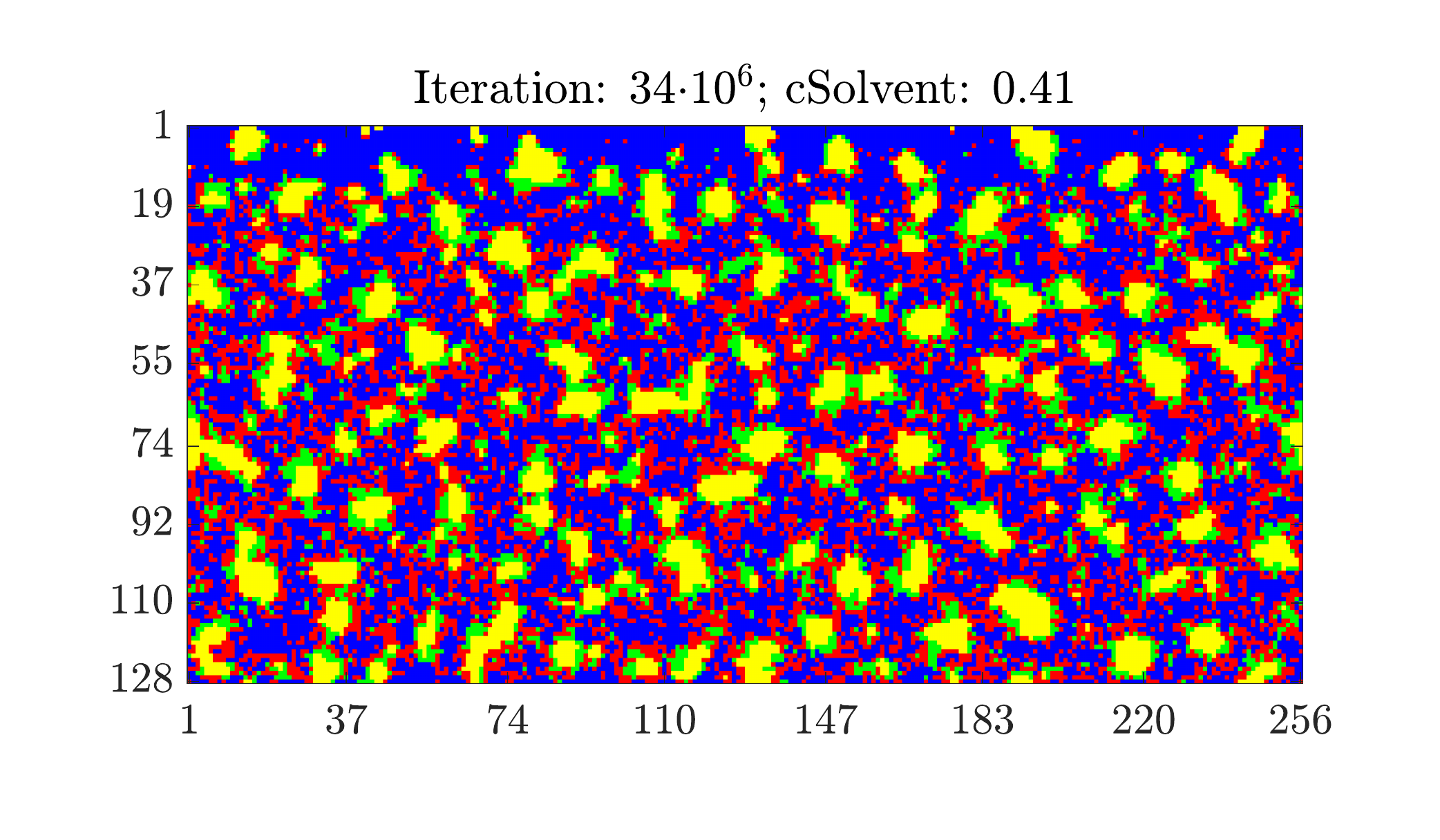}%
    \includegraphics[width=0.33\linewidth]{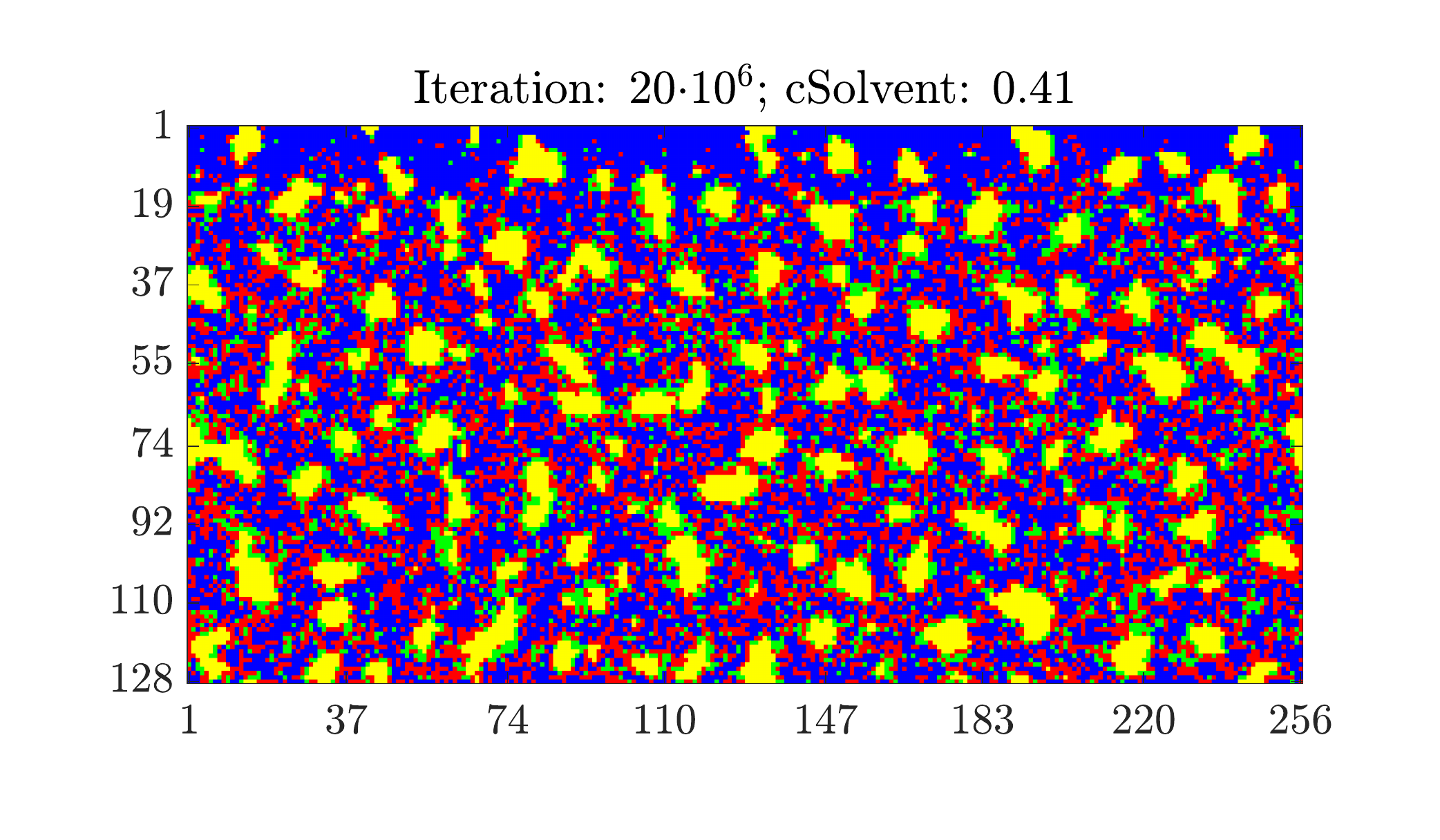}%
    \includegraphics[width=0.33\linewidth]{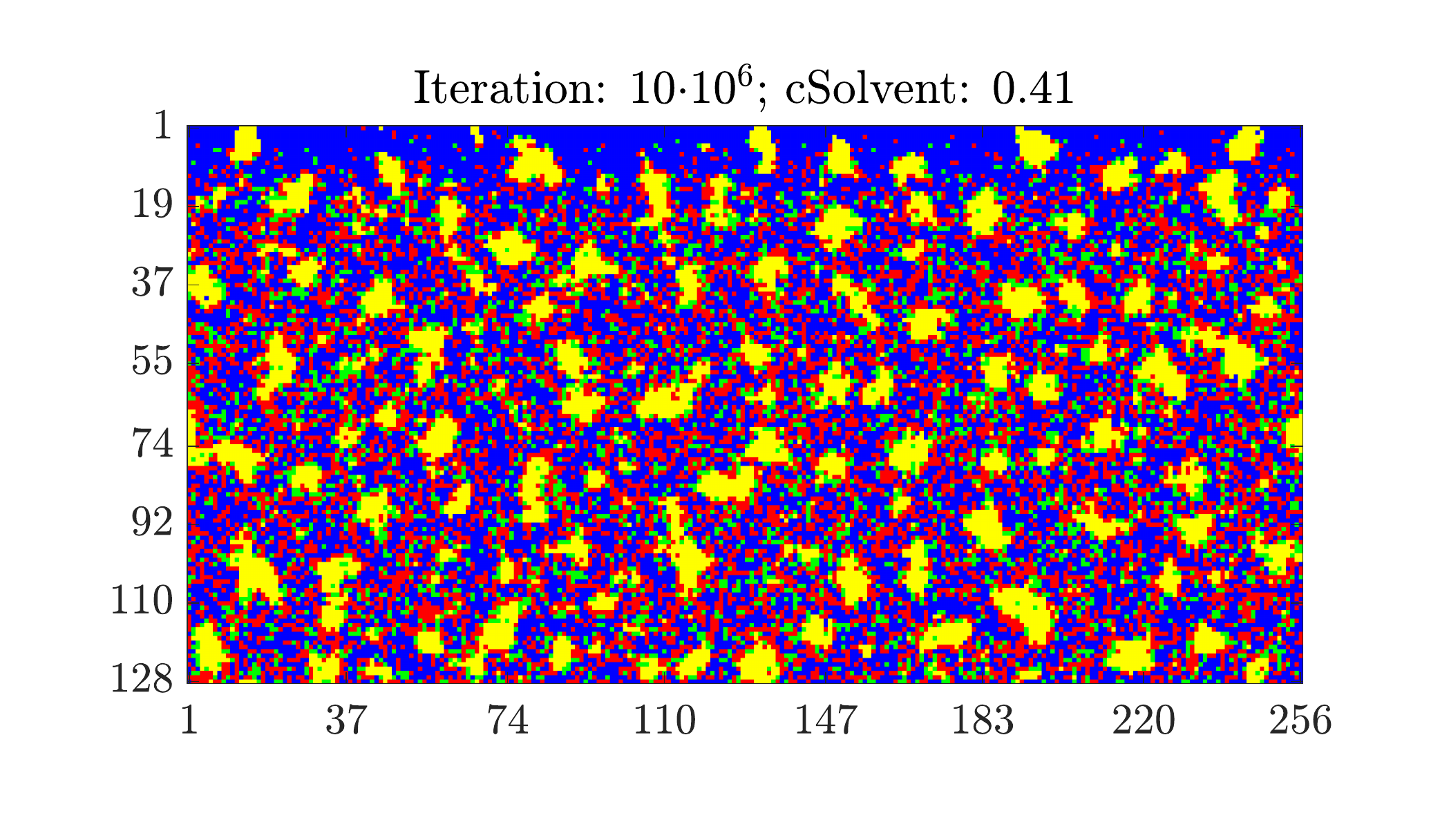}\\
    \includegraphics[width=0.33\linewidth]{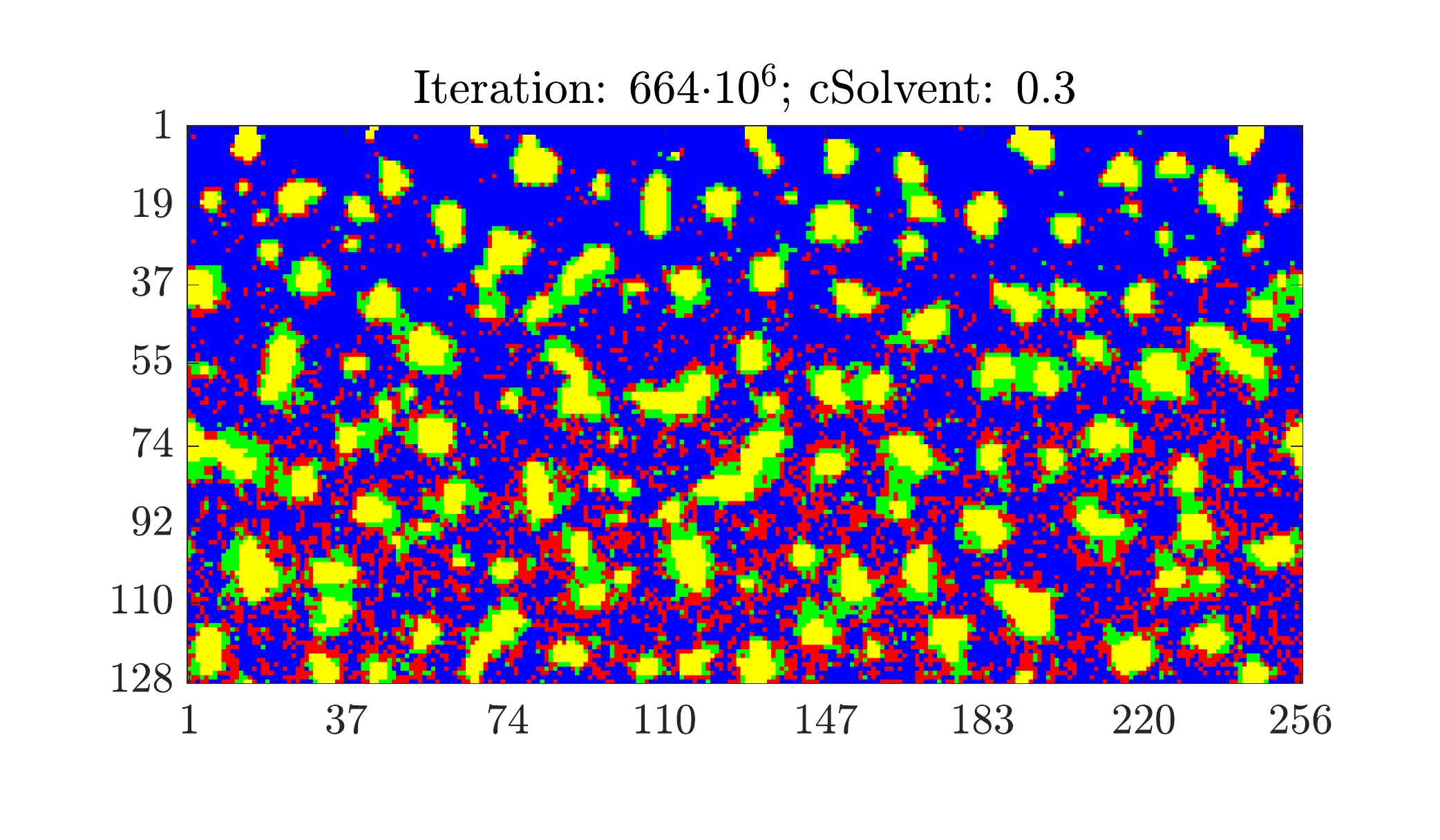}%
    \includegraphics[width=0.33\linewidth]{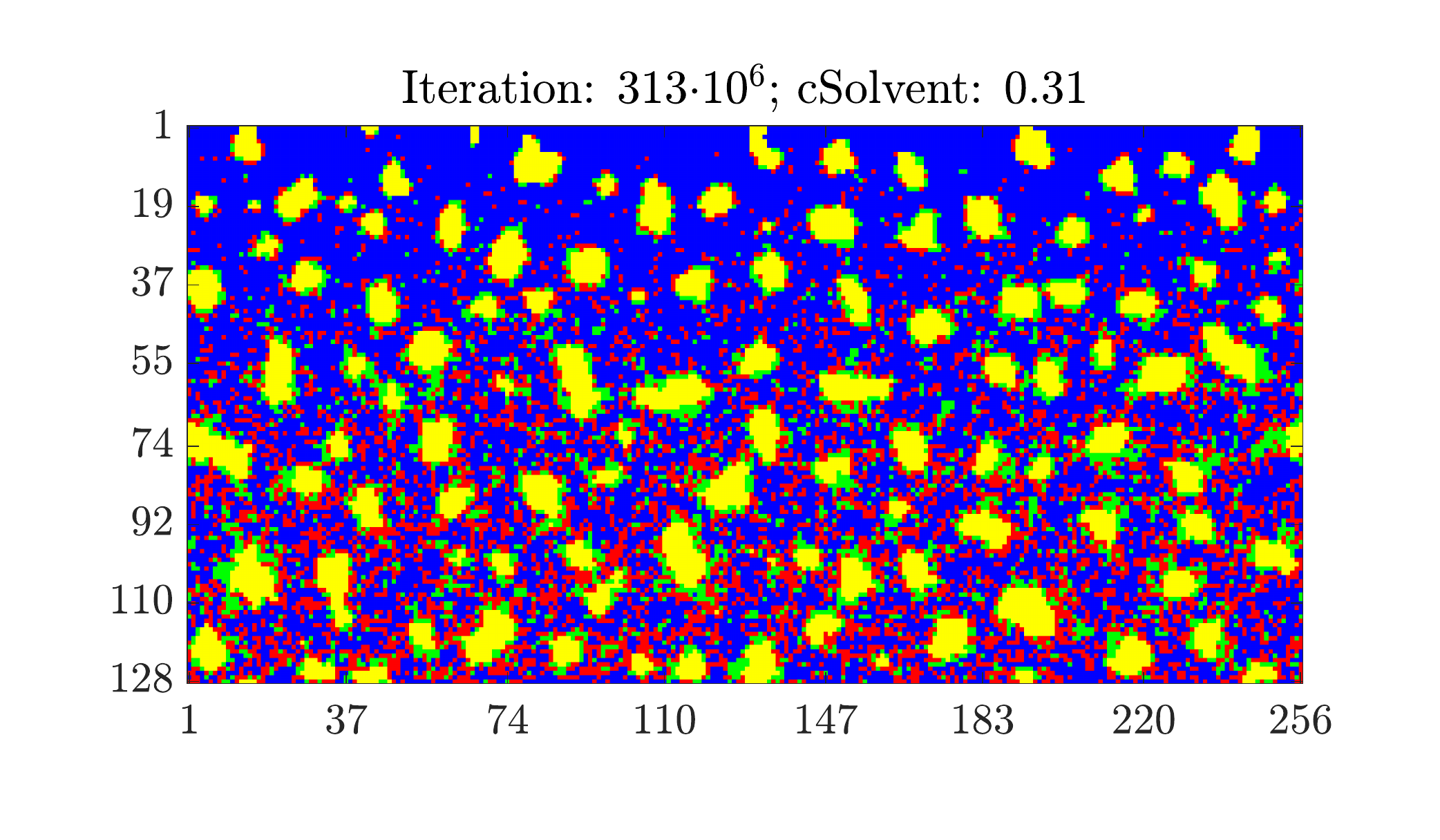}%
    \includegraphics[width=0.33\linewidth]{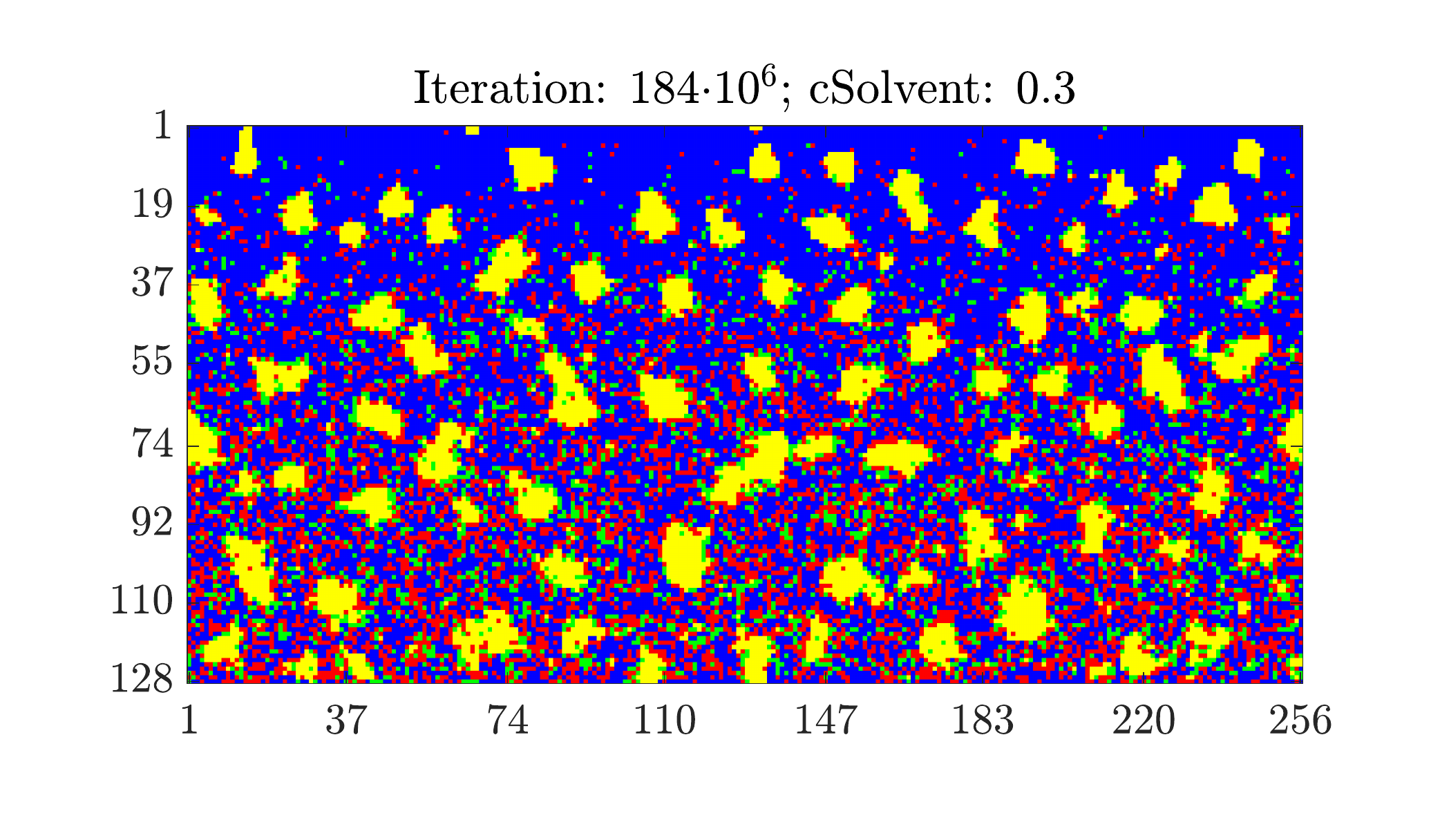}\\
    \includegraphics[width=0.33\linewidth]{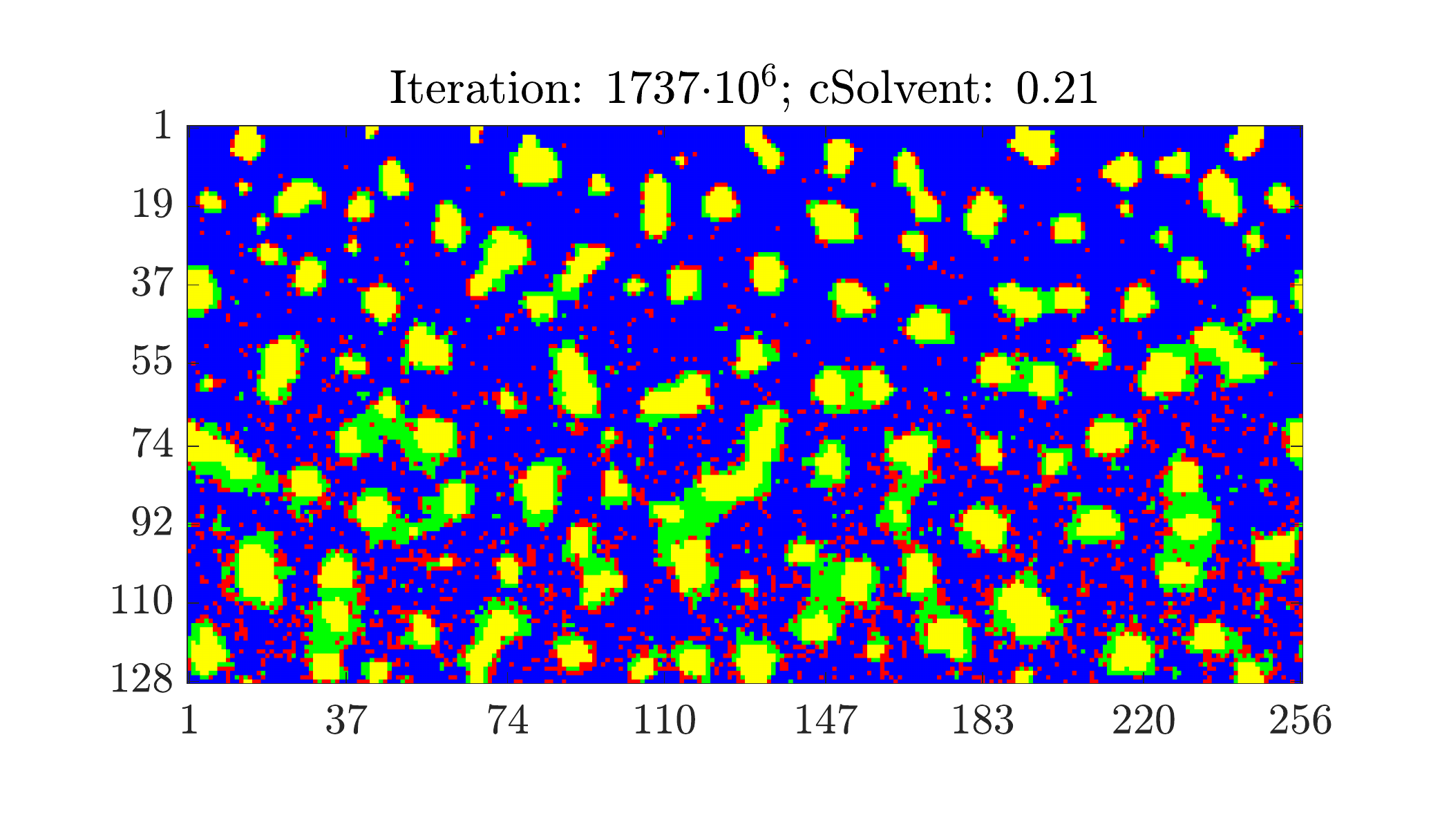}%
    \includegraphics[width=0.33\linewidth]{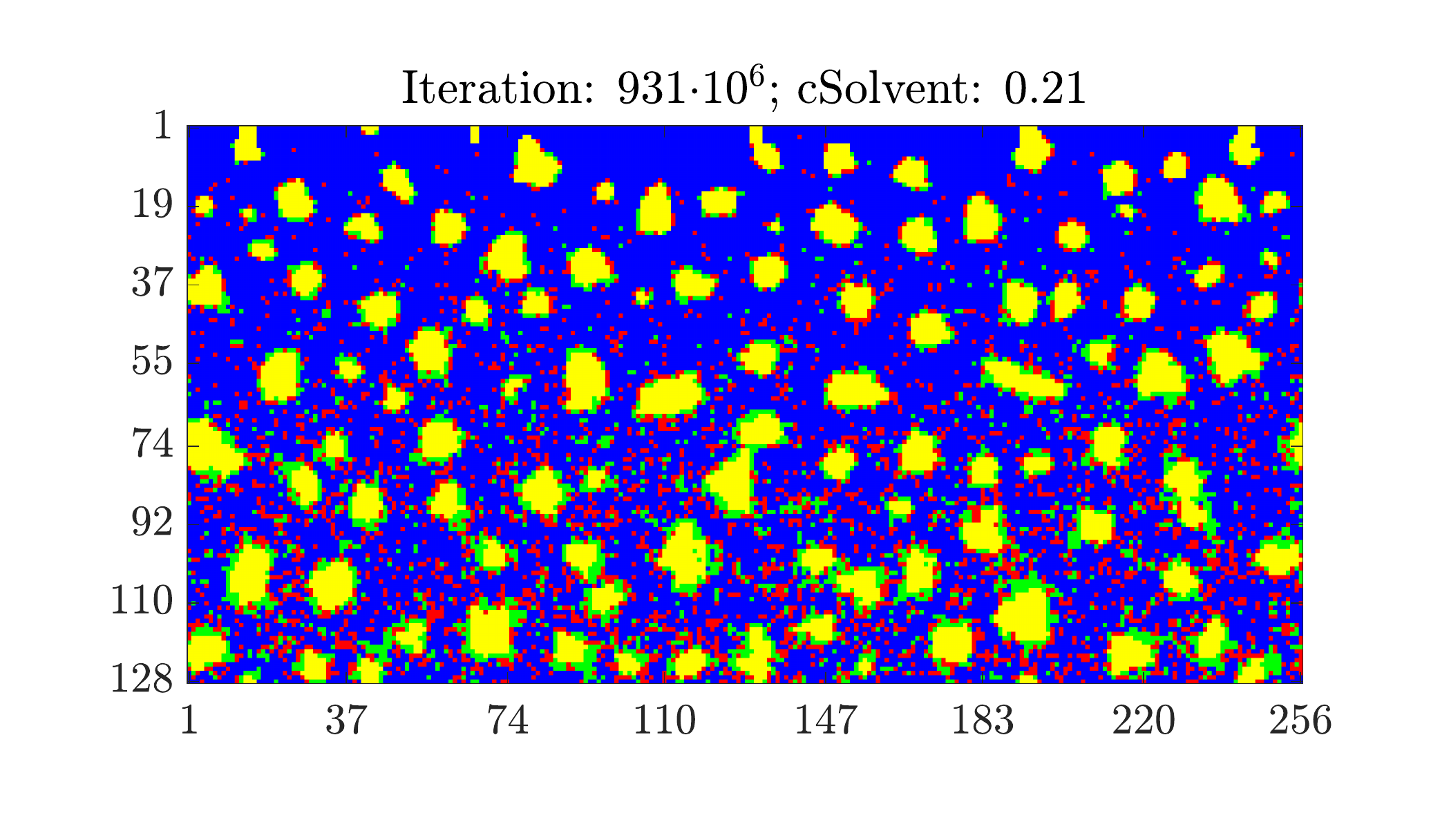}%
    \includegraphics[width=0.33\linewidth]{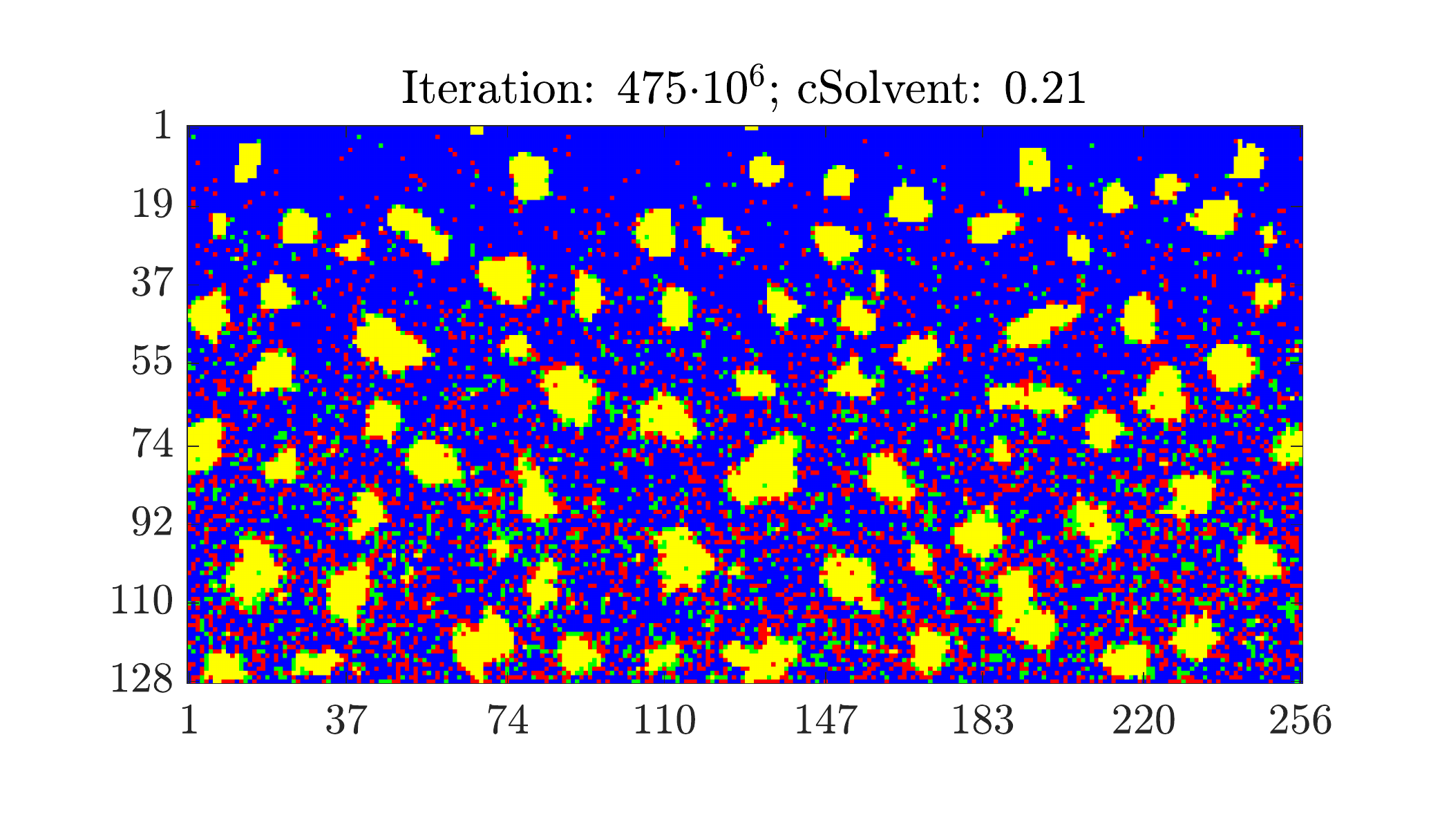}\\
    \includegraphics[width=0.33\linewidth]{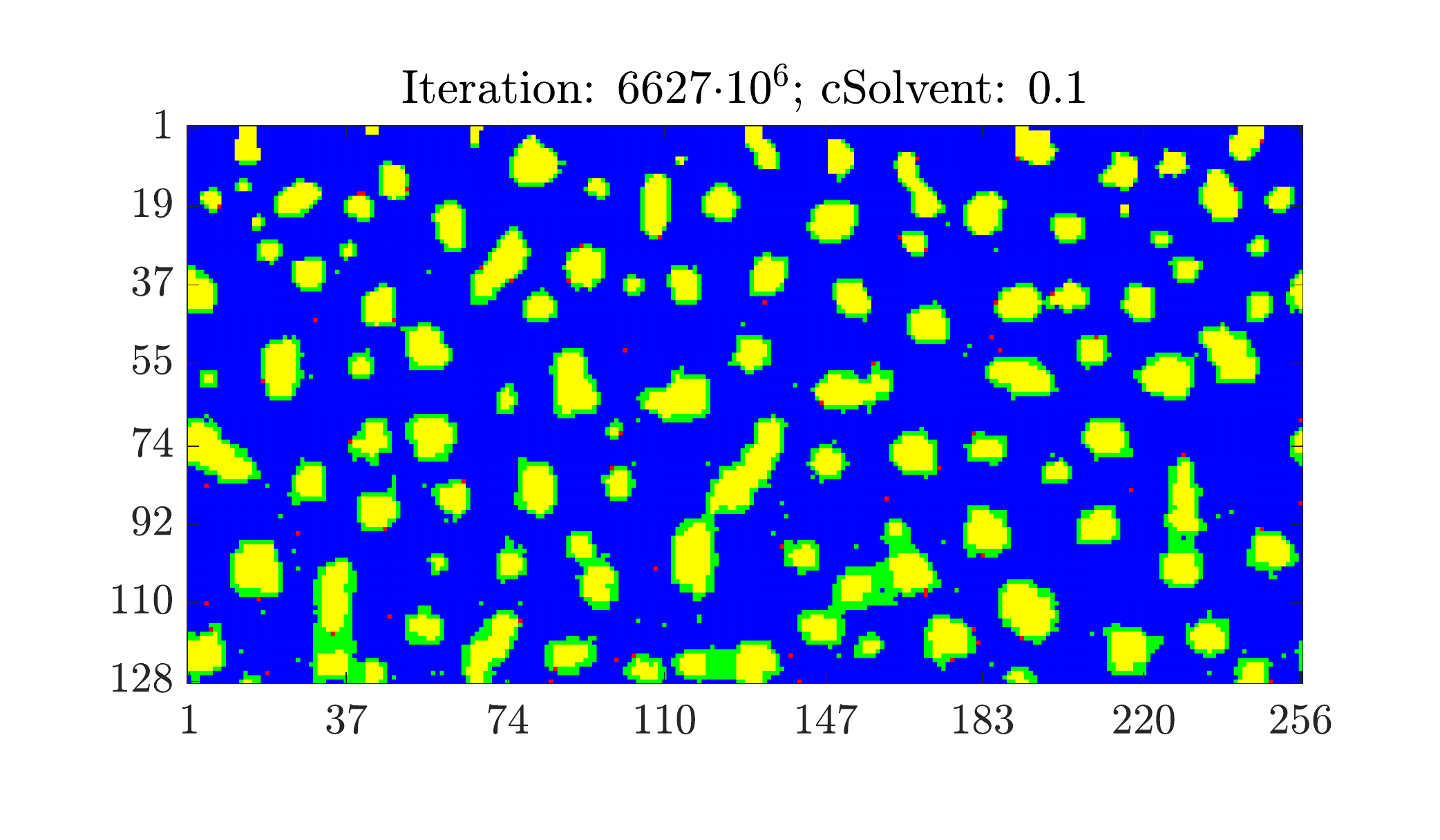}%
    \includegraphics[width=0.33\linewidth]{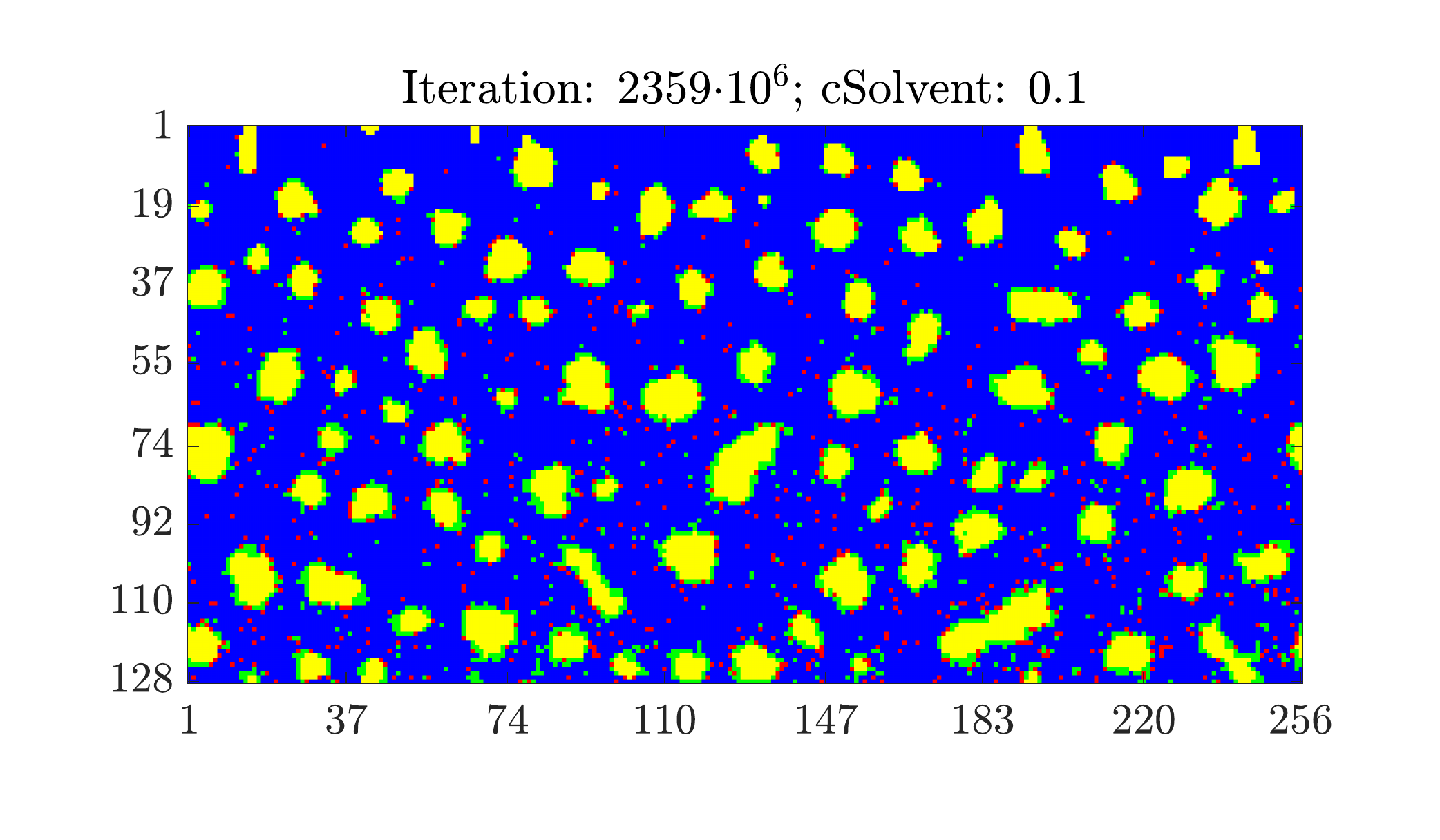}%
    \includegraphics[width=0.33\linewidth]{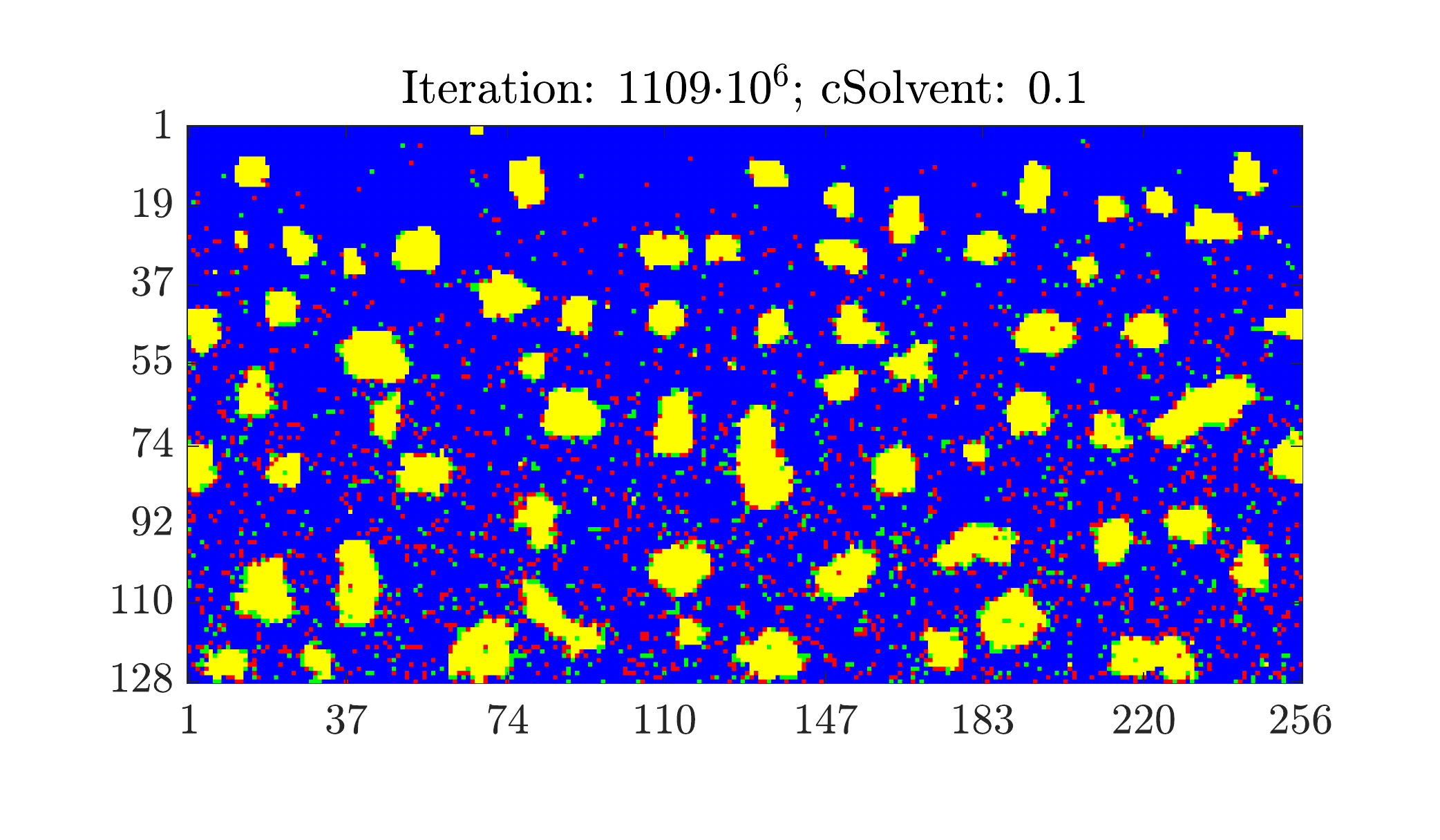}\\
    \includegraphics[width=0.33\linewidth]{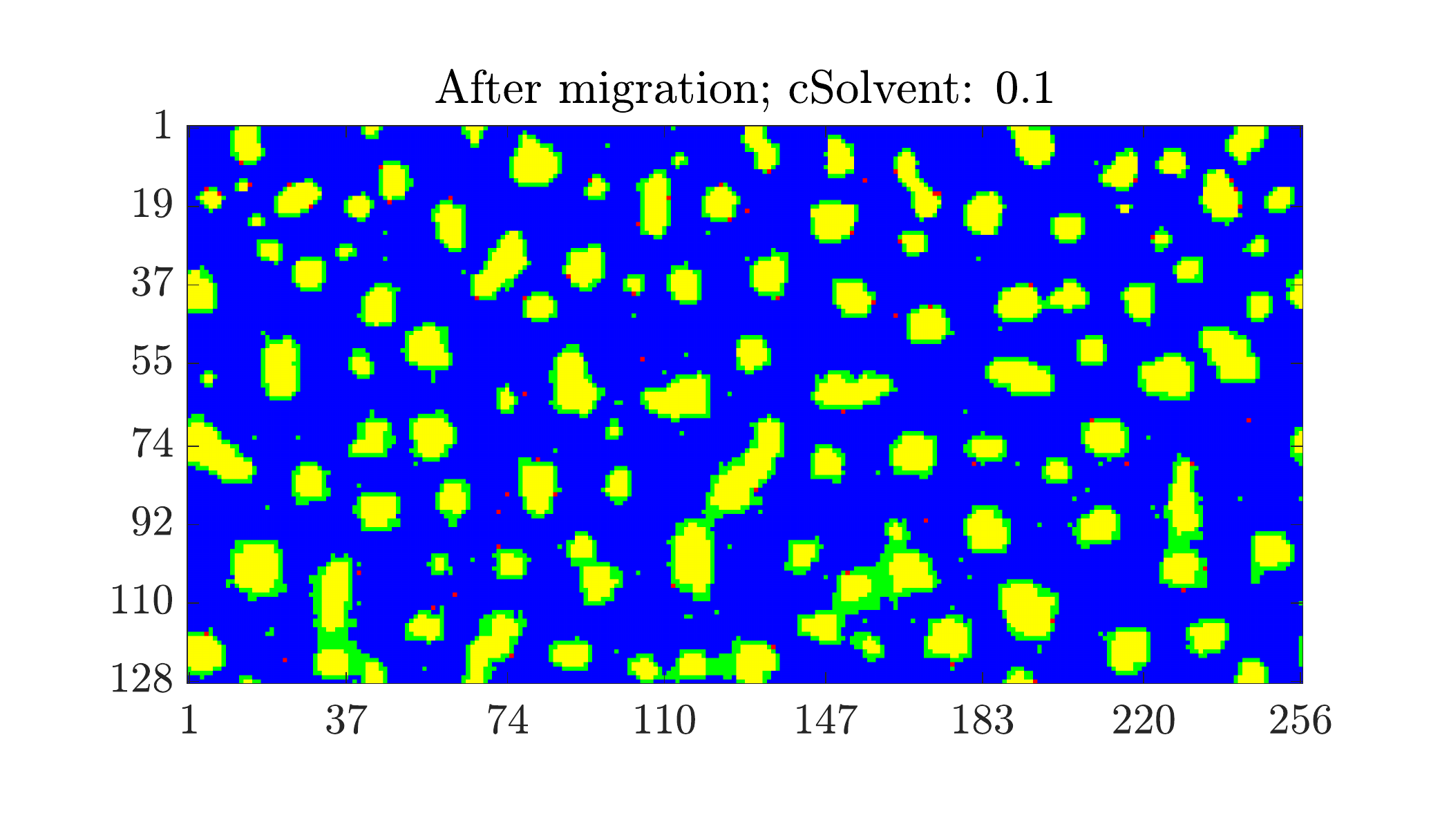}%
    \includegraphics[width=0.33\linewidth]{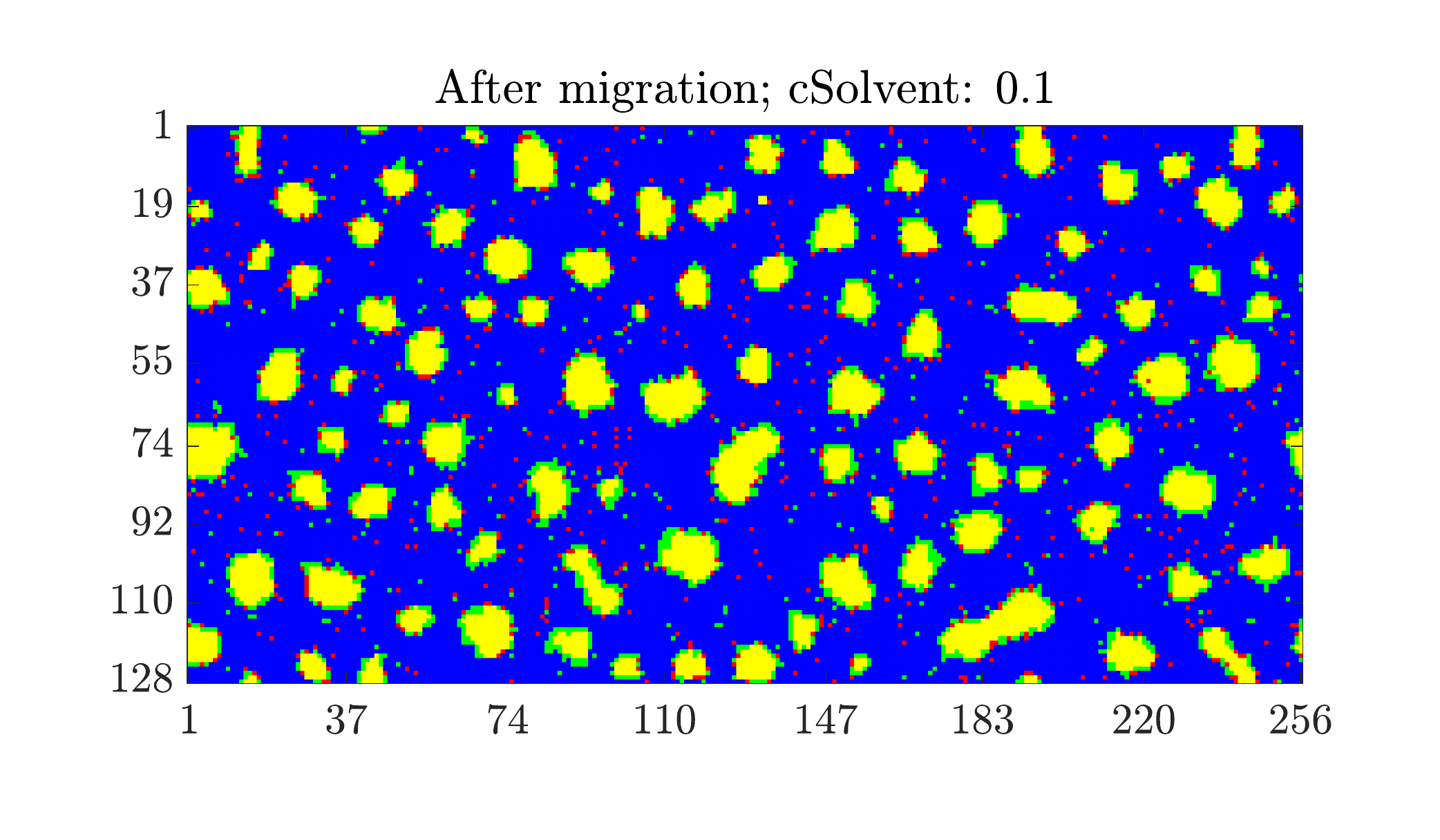}%
    \includegraphics[width=0.33\linewidth]{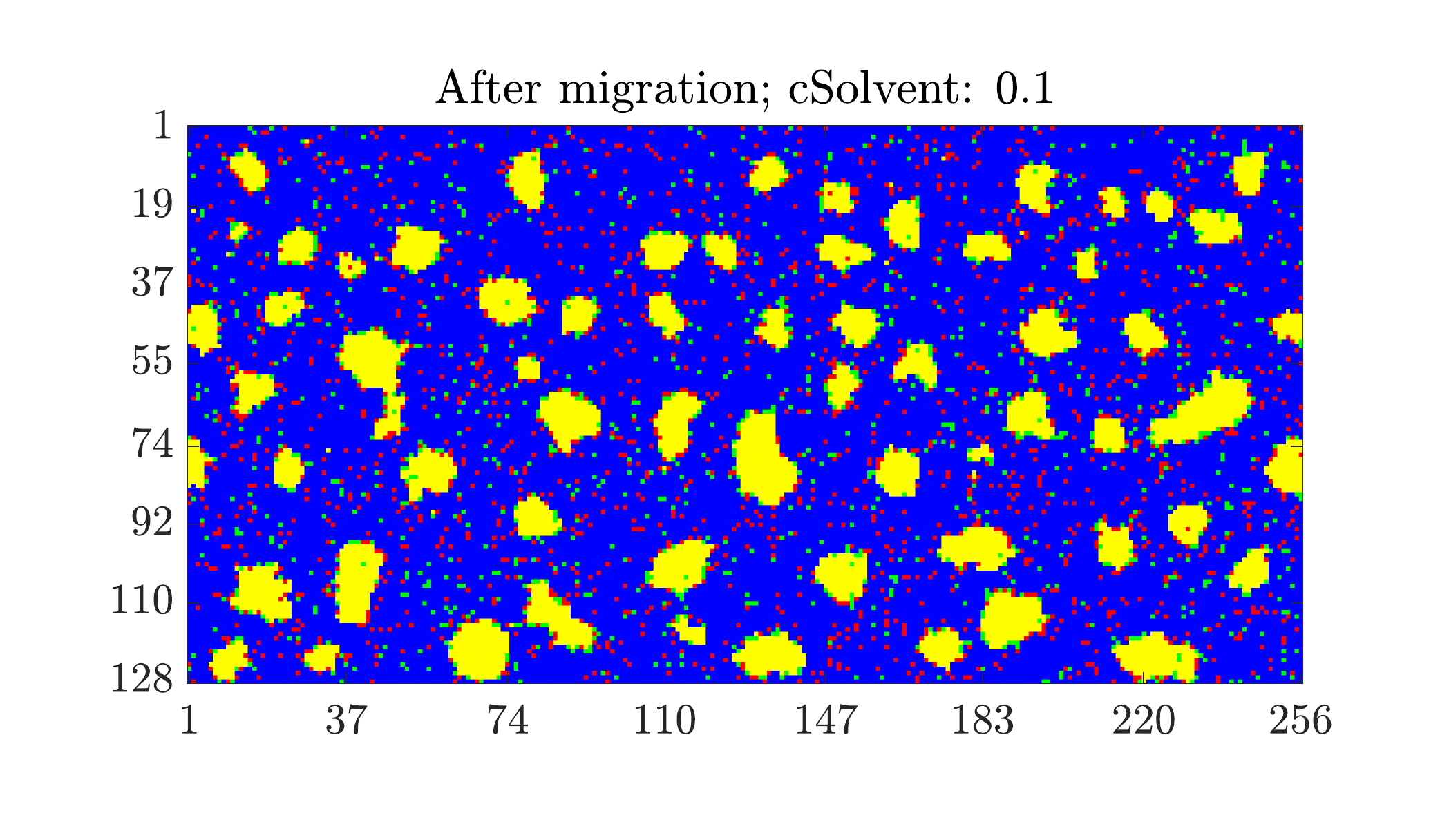}\\
    \caption{The effect of changing the temperature. Note that increasing temperature is related to more domain growth and less condensation of the solvents.}
    \label{fig:temperature}
\end{figure}

\subsection{The effect of varying temperature}

The next effect we have explored is that of varying the temperature in the system, i.e., changing the parameter $\beta$, directly proportional to the inverse temperature. 
Since we study a small section of a larger sample, the temperature within our domain will always be held constant.
In the experimental setup, however, one often observes a temperature gradient; see \cite{Macromolecules} for a related setting. 
Whence, altering the temperature of our system could be viewed as observing different slices along the temperature gradient in the experimental setup.

The simulations are shown in Figure~\ref{fig:temperature}, where the common initial configuration is shown in the top row.
As before, successive rows show the systems at 10\% reduction of solvent in the system, and the dynamics are once more switched off at 10\% remaining solvent (second to last row).
This time, each column represents different temperatures, going from cold (first column; $\beta = 0.9$) to warm (third column; $\beta = 0.3$) with an intermediate temperature of $\beta = 0.6$ in the second column.

Increasing temperature does not seem to have a drastic effect on the domain growth, but it related to less condensation of the solvents.
For example, in the first column (coldest system), the green solvent condenses around the rubber structures (yellow), and they stay condensed throughout the evolution of the system.
Returning to the interaction matrix Eq.~\eqref{eq:JMatrix}, this behaviour is easily understandable, since this solvent is less repelled by the yellow sites compared to the blue ones.
The low temperature means that non-energetically favourable movements are significantly less likely to occur, so that the condensate is stable.

In the second column, the green solvent still condenses around the yellow phase, but it does not produce ``solvent bridges'' between isolated yellow regions, like in the coldest case.
Instead, the yellow regions connected via such solvent bridges before, now often merge to one larger yellow structure.

Finally, for the warmest system in the third column, the green solvent no longer readily condenses around the yellow phase and instead evaporates together with the red solvent.
Note that the red solvent evaporates more easily in general, since it is less repulsed by the blue phase, and hence is able to migrate to the top boundary, even at lower temperatures.

\section{Conclusion and outlook}\label{Outlook}

We conclude our work with a few thoughts what concern further possible investigations in the same direction, viz. 
\begin{itemize}
\item The model captures the diffusion of the many components of the mixture, their interactions, as well as the evaporation mechanism of the solvents. The obtained morphologies are in the expected physical range. 
\item 
The interplay between the temperature and the two solvents is very complex. Interestingly, looking at Figure \ref{fig:temperature}, we can see that solvent 1 (the red solvent) can diffuse to the top boundary more easily, hence it can evaporate more rapidly. On the other hand, solvent 2 (the green solvent) tends to stay surrounding the rubber morphologies. If enough energy is available (i.e., at low values of $\beta$, associated with high temperatures), then also this solvent can diffuse to the top boundary and then evaporate.
\item The disks formation phase and the migration (terminal) phase can potentially be used for the purpose of morphology design. This would natural involve not only a rather involved optimization step but also more information on the physics and chemistry of the involved components in the mixture.
\item As further study, we think it is worth estimating numerically the coarsening rates of the rubber balls and seeing how do they depend on the solvent-solute interaction parameters. Such study would need to involve a careful quantitative analysis of the obtained morphologies in terms of correlation and structure factors calculations; the procedure set up in \cite{Andrea_PhysRevE} can be adapted to be applicable to the scenario presented here.
\end{itemize}

\section{Acknowledgments}
 An initial formulation of the problem was posed as scientific challenge to MiMM Day\textsuperscript{\textregistered} 2021 (Mathematics with Industry Day) by researchers from the company tesa SE (Germany). We thank our collaborators E. N. M. Cirillo (La Sapienza Univ., Rome, IT) and M. Colangeli (Univ. of L'Aquila, IT) for what we have learnt from them during the years concerning lattice-based modeling and simulation. 

\bibliographystyle{alpha}
\newcommand{\etalchar}[1]{$^{#1}$}

\end{document}